\documentclass[letterpaper,11pt]{article}
\pdfoutput=1 

\usepackage{jcappub} 

\usepackage[T1]{fontenc} 

\def\beq{\begin{equation}}
\def\eeq{\end{equation}}

\DeclareMathOperator{\sech}{sech}
\DeclareMathOperator{\csch}{csch}

\DeclareMathOperator{\arcsinh}{arcsinh}

\title{Perturbative analysis of the reheating dynamics of $\alpha$-attractors}

\author[]{Gabriel Germ\'an}

\affiliation[]{Instituto de Ciencias F\'{\i}sicas, Universidad Nacional
Aut\'onoma de M\'exico,\\ Av. Universidad s/n, Cuernavaca, Morelos, 62210, Mexico}

\emailAdd{gabriel@icf.unam.mx}

\abstract{We study the reheating phase following inflation in the context of single-field models, focusing on the perturbative decay of the inflaton into lighter particles. A general analytical framework is presented to compute the reheating temperature $T_{re}$ and related quantities by combining cosmological observations with model-dependent parameters. We derive expressions for $T_{re}$ for three types of interactions: gravitational, scalar, and Yukawa-type fermionic couplings, and apply these results to the class of $\alpha$-attractor inflationary models, which exhibit attractor behavior in the $(n_s, r)$ plane. The main goal of this work is to investigate how key cosmological quantities such as $T_{re}$, $N_{re}$, and $m_\phi$ among others, evolve with the scalar spectral index $n_s$ and the Yukawa coupling constant $y$, within a consistent analytical framework. Although the formulas used are approximate, they are sufficient to capture the qualitative behavior of the relevant quantities across a wide range of parameter values. Here, we are not interested in precise numerical approximations or data analysis, but rather in understanding the general trends and dependence of cosmological quantities of interest. In particular, tendencies observed in the figures, such as the sensitivity of $T_{re}$ to the coupling strength and the equation-of-state parameter $\omega_{re}$, reflect physical features that are not strongly affected by the approximations involved. }

\begin{document}
\maketitle
\flushbottom

\section{Introduction} \label{Intro}
Reheating is the process that connects the end of inflation to the beginning of the radiation-dominated era. It is characterized by three key quantities: the equation-of-state parameter $\omega_{re}$, the reheating temperature $T_{re}$, and the duration of reheating, typically expressed in terms of the number of $e$-folds $N_{re}$ \cite{Bassett:2005xm}-\cite{Amin:2014eta}. These parameters play a central role in determining the thermal history of the universe, influencing particle production, thermalization, and leaving imprints on cosmological observables such as the primordial power spectrum and the cosmic microwave background (CMB). While the perturbative decay of the inflaton into lighter particles provides a tractable framework for studying reheating, it has important limitations. This approach assumes that the inflaton decay can be treated as a sequence of individual particle interactions, which may break down when collective effects become significant. In particular, the perturbative treatment neglects
resonant particle production during the coherent oscillations of the inflaton field, back-reaction effects from the created particles on the inflaton dynamics, non-linear interactions that can lead to thermalization and possible non-equilibrium effects during the energy transfer process.
Despite these limitations, the perturbative framework remains valuable for understanding the general features of reheating and establishing connections between inflationary models and observable consequences. 
By combining observational CMB data with inflationary models (for reviews on inflation see e.g., \cite{Linde:1984ir}-\cite{Martin:2013tda}), it is possible to place constraints on the reheating phase, which in turn translate into bounds on inflationary observables such as the scalar spectral index $n_s$ and the tensor-to-scalar ratio $r$. This connection provides a valuable consistency check for inflationary scenarios and can provide viable ranges for model parameters.

The aim of this work is to explore the qualitative behavior of reheating-related quantities in single-field inflation models with perturbative inflaton decay. We make use of analytical expressions that, while approximate, are sufficient to reveal how quantities like $T_{re}$, $N_{re}$, and $m_\phi$ among others, vary with the spectral index $n_s$ and the (Yukawa) coupling constant $y$. The focus is not on obtaining precise numerical values or to perform data analysis, but rather on understanding the general trends and dependencies visible in the figures. We discuss three types of interactions: gravitational, scalar, and Yukawa-type fermionic couplings. For generality, we assume an inflationary potential of the form $V(\phi) = V_0 f(\phi)$, where $f(\phi)$ contains the model dependence. Once the inflaton field value at horizon crossing $\phi_k$, is determined, the entire set of relevant quantities can be computed consistently. This formalism is applied to $\alpha$-attractor models which are characterized by a parameter $\alpha$ that controls the geometry of the scalar field space and the shape of the potential. $\alpha$-attractors predict a narrow range of values for $n_s$ and $r$, consistent with current observational bounds \cite{Ferrara:2013rsa}-\cite{Kallosh:2013tua}. Our study provides insight into how cosmological quantities depend on both the parameters of the model and the inflationary observables, offering a picture of the dynamics of the early universe in this class of scenarios.

The article is organized as follows: In Section~2, we briefly review the theoretical background linking inflationary dynamics to the reheating phase. We discuss the perturbative decay of the inflaton field and its role in determining the reheating temperature $T_{re}$, presenting a general expression for $T_{re}$ and simple formulas for the inflaton mass and for the maximum temperature attainable in the limit of instantaneous reheating. This formula can be used to impose an upper bound to the inflaton coupling. In Section~3, we consider three types of inflaton interactions: gravitational, scalar, and Yukawa-type fermionic couplings. For each case, we provide the relevant decay rates and corresponding expressions for the reheating temperature. Section~4 presents a detailed numerical analysis of the $\alpha$-attractor inflationary model. We include plots of the reheating temperature, the number of $e$-folds during inflation, reheating, and radiation, as well as the scalar spectral index $n_s$, its running $n_{sk}$, and the tensor-to-scalar ratio $r$. These quantities are shown as functions of $n_s$ and the Yukawa coupling $y$, for a range of values of the equation-of-state parameter $\omega_{re}$. In Section~5, we briefly discuss non-perturbative reheating effects, such as preheating, and their potential impact on the evolution of the early universe. Finally, Section~6 summarizes our main results and conclusions.
\section{General Framework}\label{gen}
A general relation connecting the inflationary and reheating phases is given by\footnote{An equivalent form of Eq.~(\ref{Nk1}) can be found in several works; see, e.g., \cite{Liddle:1994dx}-\cite{German:2023yer}. For a derivation leading exactly to this form, see  \cite{German:2024rmn}. Note that the third term in Eq.~(\ref{Nk1}) vanishes for $\omega_{re}=1/3$ apparently  implying no dependence on $T_{re}$ for this case. However, Eq.~(\ref{Nk1}) is an equation for $\phi_k$, the value of the inflaton at horizon crossing, not for $T_{re}$. For the inflaton decaying into fermions, as will be studied below, $T_{re}$ is given by Eq.~(\ref{Trey2}). Thus, for $\omega_{re}=1/3$, the vanishing of third term simply implies that it does not contribute to the determination of $\phi_k$.}
\begin{align}
\label{Nk1}
N_k = & \ln\left( \frac{2}{k_p} \left( \frac{43}{11 g_{s,re}} \right)^{1/3} \pi \sqrt{A_s} a_0 T_0 \right) 
+ \frac{1}{3(1 + \omega_{re})} \ln\left( \frac{g_{re}}{540 A_s} \right) \nonumber 
+ \frac{1 - 3 \omega_{re}}{3(1 + \omega_{re})} \ln \left( \frac{T_{re}}{M_{Pl}} \right)\\
&+ \frac{1}{3(1 + \omega_{re})} \ln\left( \frac{\left( M_{Pl} f^{\prime}(\phi_k) \right)^{1 + 3 \omega_{re}}}{f(\phi_k)^{3 \omega_{re}} f(\phi_{end})} \right),
\end{align}
where $N_k \equiv \ln(a_e/a_k)$ is the number of $e$-folds of inflation after the pivot scale $k_p = 0.05\, \mathrm{Mpc}^{-1}$ exits the horizon. Here, $A_s$ is the amplitude of the scalar power spectrum, $a_0$ and $T_0$ are the present-day scale factor and CMB temperature,  and $g_{re},  g_{s,re}$ are the effective relativistic and entropy degrees of freedom at the end of reheating, respectively. The effective equation-of-state parameter during reheating is denoted $\omega_{re}$, and $T_{re}$ is the reheating temperature. The inflationary potential is assumed to take the form $V(\phi) = V_0 f(\phi)$, where $f(\phi)$ captures the model-dependent structure. The derivative $f^{\prime}(\phi_k)$ is evaluated at the field value $\phi_k$ when the pivot scale exits the horizon.

Equation~(\ref{Nk1}) consists of a constant term depending on cosmological parameters, terms involving $\omega_{re}$ and $T_{re}$, and a model-dependent term involving $f(\phi)$ and $f^{\prime}(\phi)$. As we will show, $T_{re}$ can be expressed in terms of the model parameter $\alpha$, the inflaton coupling $y$, and the field value $\phi_k$. Consequently, we treat $(\alpha, y, \omega_{re})$ as free parameters. For a fixed model and given triplet $(\alpha, y, \omega_{re})$, we solve Eq.~(\ref{Nk1}) to determine $\phi_k$, from which all relevant cosmological quantities can be computed.

To derive analytical expressions for $T_{re}$, the inflaton mass $m_\phi$, and the maximum reheating temperature, we proceed as follows.
At the end of reheating, the inflaton energy density is approximately related to the decay rate by
\begin{equation}
\label{rore}
\rho_{re} \approx 3 H_{re}^2 M_{Pl}^2 \approx 3 \Gamma_\phi^2 M_{Pl}^2.
\end{equation}
Reheating ends when the energy densities of radiation and the inflaton become comparable. The radiation energy density is given by
\begin{equation}
\label{rorad}
\rho_{re} \approx \rho_{rad} = \frac{\pi^2 g_{re}}{30} T_{re}^4.
\end{equation}
Equating Eqs.~(\ref{rore}) and (\ref{rorad}) yields
\begin{equation}
\label{Tre}
T_{re} \approx \left( \frac{90 \, \Gamma_\phi^2}{g_{re} \pi^2 M_{Pl}^2} \right)^{1/4} M_{Pl}.
\end{equation}
The inflaton mass is defined via the second derivative of the potential at its minimum
\begin{equation}
\label{mass1}
m_\phi^2 = V^{\prime\prime}(\phi_0) \equiv \gamma \frac{V_0}{M_{Pl}^2},
\end{equation}
where $\gamma$ is a dimensionless model-dependent parameter.
The scalar amplitude at horizon crossing is
\begin{equation}
\label{A1}
A_s(k) = \frac{1}{24\pi^2} \frac{V(\phi_k)}{\epsilon_k M_{Pl}^4}.
\end{equation}
Using $V(\phi_k) = V_0 f(\phi_k)$, we can solve for $V_0$
\begin{equation}
\label{V0}
V_0 = \frac{12 A_s \pi^2 f^{\prime}(\phi_k)^2 M_{Pl}^6}{f(\phi_k)^3}.
\end{equation}
The inflaton mass then becomes
\begin{equation}
\label{mass2}
m_\phi \approx \sqrt{\gamma} \left( \frac{12 A_s \pi^2 f^{\prime}(\phi_k)^2 M_{Pl}^2}{f(\phi_k)^3} \right)^{1/2} M_{Pl}.
\end{equation}
Finally, the number of $e$-folds during reheating is given by
\begin{equation}
\label{Nre}
N_{re} = \frac{1}{3(1 + \omega_{re})} \ln\left( \frac{\rho_{end}}{\rho_{re}} \right).
\end{equation}
The condition $N_{re} = 0$ corresponds to instantaneous reheating, $\rho_{end} = \rho_{re}$, where\footnote{From the second Friedmann equation at the end of inflation $\ddot a=0$. Thus, $\rho_{end}=-3p_{end}$ or $\dot\phi_{end}^2=V_{end}$, from where the result (\ref{roend}) follows.}
\begin{equation}
\label{roend}
\rho_{end} \approx \frac{3}{2} V_{end} = \frac{3}{2} V_0 f(\phi_e).
\end{equation}
Substituting into Eq.~(\ref{Tre}) gives the maximum reheating temperature
\begin{equation}
\label{Tmax}
T_{max} \approx \left( \frac{540 A_s f(\phi_e) f^{\prime}(\phi_k)^2 M_{Pl}^2}{f(\phi_k)^3 g_{re}} \right)^{1/4} M_{Pl}.
\end{equation}
This upper limit provides a theoretical bound on the decay coupling in a given model. These general expressions apply to any single-field inflationary scenario for specific decay mechanisms.
\section{Inflaton Decay Mechanisms}\label{decay}
We now consider three decay channels for the inflaton field $\phi$: gravitational interactions, scalar interactions, and Yukawa-type fermionic interactions. The decay rate $\Gamma_\phi$ determines the reheating temperature and depends on the nature of the coupling.
\subsection{Gravitational Decay}\label{gravitational}
The inflaton can couple gravitationally to other fields via non-minimal couplings to the energy-momentum tensor. The leading order interaction with a light scalar field $\chi$ is given by
\begin{equation}
\label{Dgra_0}
\mathcal{L}_{\mathrm{int}} = \frac{1}{2 M_{Pl}} \phi\, T_\chi,
\end{equation}
where $T_\chi$ is the trace of the energy-momentum tensor of $\chi$. For a massless scalar, $T_\chi = 0$ due to conformal invariance. If $\chi$ has mass $m_\chi$, then
\begin{equation}
T_\chi = g^{\mu\nu} T_{\mu\nu} = - m_\chi^2 \chi^2
\end{equation}
and the interaction becomes
\begin{equation}
\label{Dgra_1}
\mathcal{L}_{\mathrm{int}} = - \frac{1}{2 M_{Pl}} m_\chi^2 \phi \chi^2.
\end{equation}
This interaction vanishes in the massless limit. However, derivative couplings can induce decay even for $m_\chi = 0$, for example
\begin{equation}
\label{Dgra_2}
\mathcal{L}_{\mathrm{int}} = \frac{\phi}{2 M_{Pl}} \partial^\mu \chi \partial_\mu \chi.
\end{equation}
This leads to a suppressed decay rate
\begin{equation}
\label{Dgra}
\Gamma_\phi \approx \frac{m_\phi^3}{192 \pi M_{Pl}^2}.
\end{equation}
This decay is inefficient due to Planck suppression. In the Starobinsky model, the inflaton (scalaron) arises from $R + \alpha R^2$ gravity and decays only via gravitational interactions, resulting in low reheating efficiency. Using Eq.~(\ref{Dgra}) and the earlier expressions, we obtain the reheating temperature for gravitational decay
\begin{equation}
\label{Tregra}
T_{re}^{\mathrm{(grav)}} \approx \left( \frac{135 A_s^3 \pi^2 f'(\phi_k)^6 M_{Pl}^{6} \gamma^3}{32 g_{re} f(\phi_k)^9} \right)^{1/4} M_{Pl}.
\end{equation}
The implications of Eq.~(\ref{Tregra}) for the $\alpha$-Starobinsky and $\alpha$-attractor models of inflation were studied in \cite{German:2024rmn} and are not discussed further here.
\subsection{Decay into Scalars}\label{scalars}
Scalar coupling interactions of the form $g \phi \chi^2$ arise naturally in models where $\phi$ is Higgs like. The interaction Lagrangian is
\begin{equation}
\label{L2}
\mathcal{L}_{\mathrm{int}} = -g \phi \chi^2,
\end{equation}
with decay rate
\begin{equation}
\label{Dga}
\Gamma_{\phi \to \chi\chi} = \frac{g^2}{8\pi m_\phi} \left( 1 - \frac{4m_\chi^2}{m_\phi^2} \right)^{1/2},
\end{equation}
which simplifies for $m_\chi \ll m_\phi$
\begin{equation}
\label{Dgb}
\Gamma_{\phi \to \chi\chi} \approx \frac{g^2}{8\pi m_\phi}.
\end{equation}
The corresponding reheating temperature becomes
\begin{equation}
\label{Treg}
T_{re}^{(g)} \approx |\tilde{g}| \left( \frac{15 f(\phi_k)^3}{128 A_s \pi^6 f'(\phi_k)^2  M_{Pl}^2 g_{re}\gamma} \right)^{1/4} M_{Pl},
\end{equation}
where $\tilde g\equiv g/M_{Pl}$. This scenario yields more efficient reheating than gravitational decay. However, due to the similarity in qualitative behavior with fermionic couplings, we do not analyze this case further.
\subsection{Decay into Fermions}\label{fermions}
If the inflaton couples to a fermionic field $\psi$ via a Yukawa interaction, the Lagrangian is
\begin{equation}
\label{Ly}
\mathcal{L}_{\mathrm{int}} = - y \phi \bar{\psi} \psi,
\end{equation}
with decay rate
\begin{equation}
\label{Dya}
\Gamma_{\phi \to \bar{\psi} \psi} = \frac{y^2 m_\phi}{8\pi} \left( 1 - \frac{4m_\psi^2}{m_\phi^2} \right)^{3/2},
\end{equation}
which simplifies to
\begin{equation}
\label{Dyb}
\Gamma_{\phi \to \bar{\psi} \psi} \approx \frac{y^2 m_\phi}{8\pi}, \quad \text{for } m_\psi \ll m_\phi.
\end{equation}
The corresponding reheating temperature is
\begin{equation}
\label{Trey1}
T_{re}^{(y)} \approx |y| \left( \frac{135 A_s f'(\phi_k)^2 M_{Pl}^2\gamma}{8 \pi^2 g_{re} f(\phi_k)^3} \right)^{1/4} M_{Pl}.
\end{equation}
Decay into fermions is studied in detail in the following section for the $\alpha$-attractor models of inflation.
\section{Reheating in $\boldsymbol{\alpha}$-Attractor Models}\label{Attractor}
We now turn our attention to $\alpha$-attractor models, which are motivated by supergravity and conformal symmetry frameworks. These models are characterized by the following scalar potential (for a representative selection of references on $\alpha$-attractors, see \cite{Linder:2015qxa}--\cite{Alestas:2024eic}; for their implications on reheating, see \cite{Martin:2014nya}--\cite{Drewes:2023bbs})
\begin{equation}
\label{pot}
V(\phi) = V_0 \tanh^2\left( \frac{\phi}{\sqrt{6\alpha} M_{Pl}} \right),
\end{equation}
where $\alpha$ is a dimensionless parameter that controls the curvature of the scalar field manifold, and $M_{Pl}$ is the reduced Planck mass. Related analyses have been carried out by the author in Refs.~\cite{German:2023yer}, \cite{German:2024rmn}, and  \cite{German:2020cbw}. The present work differs from those in several aspects. In Ref.~\cite{German:2023yer}, the number of $e$-folds during reheating, $N_{re}$, is expressed in terms of the scalar spectral index $n_s$ and the tensor-to-scalar ratio $r$ by eliminating the model parameters $\alpha$ and $V_0$. By imposing the condition $N_{re} \geq 0$ (see Fig.~2 of Ref.~\cite{German:2023yer}, for the particular case $\omega_{re}=0$), bounds on $n_s$ and $r$ are obtained, which then constrain other cosmological parameters. Similar constraints are derived by applying bounds from Big Bang Nucleosynthesis (BBN) on the reheating temperature. These results are summarized in tables showing the corresponding bounds. A comparable analysis is presented in Ref.~\cite{German:2020cbw}, where the potential $V = V_0 \left(1 - \mathrm{sech}\left(\frac{\phi}{\sqrt{6\alpha} M_{\text{Pl}}}\right)^p \right)$ is studied for $p = 1$ and $p = 4$, in addition to the $p = 2$ case.
In contrast, the current study numerically solves the full equation~(\ref{Nk1}) for the inflaton field value at horizon crossing, $\phi_k$, and uses this to analyze the behavior of cosmological observables across a range of plausible values for the reheating equation-of-state parameter $\omega_{re}$ and selected values of $\alpha$. The results are presented as functions of the number of $e$-folds during inflation, $N_k$, the scalar spectral index $n_s$, and, in the specific case of inflaton decay into fermions, the Yukawa coupling. This coupling enters Eq.~(\ref{Nk1}) through the reheating temperature $T_{re}$ as given by Eq.~(\ref{Trey2}). Ref.~\cite{German:2024rmn}  is more similar in scope to the present work, but instead considers gravitational reheating, where the inflaton does not couple directly to matter fields (see Subsection~3.1). In the present study, the numerical approach allows for a direct visualization of how cosmological quantities vary with model parameters. The color-coded representation of $\omega_{re}$ provides an immediate indication of the relevant parameter space and its physical implications. 
  
To study inflationary dynamics, we compute the standard slow-roll parameters, defined as
\begin{align}
\epsilon &\equiv \frac{M_{Pl}^2}{2} \left( \frac{V'}{V} \right)^2, &
\eta &\equiv M_{Pl}^2 \frac{V''}{V}, &
\xi &\equiv M_{Pl}^4 \frac{V' V'''}{V^2}.
\end{align}
These parameters allow us to express the key inflationary observables within the slow-roll approximation. Specifically, the scalar spectral index $n_s$, its running $n_{sk}$ (usually denoted by $\alpha$), and the tensor-to-scalar ratio $r$ are given by
\begin{align}
n_s &= 1 + 2\eta - 6\epsilon, \label{ns} \\
n_{sk} &= \frac{d n_s}{d \ln k} = 16\epsilon\eta - 24\epsilon^2 - 2\xi, \label{nsk} \\
r &= 16\epsilon. \label{r}
\end{align}
The number of $e$-folds between horizon exit of the pivot scale and the end of inflation is
\begin{equation}
\label{Nk}
N_k = \frac{3}{2} \alpha \left[ \cosh^2\left( \frac{\phi_k}{\sqrt{6\alpha} M_{Pl}} \right) - \cosh^2\left( \frac{\phi_{end}}{\sqrt{6\alpha} M_{Pl}} \right) \right],
\end{equation}
where $\phi_k$ denotes the value of the inflaton field at horizon exit, and $\phi_{end}$ denotes its value at the end of inflation. The end of inflation is determined by the condition $\epsilon(\phi_{end}) = 1$, which leads to the following expression
\begin{equation}
\phi_{end} = \sqrt{\frac{3\alpha}{2}} M_{Pl} \arcsinh\left( \frac{2}{\sqrt{3\alpha}} \right).
\end{equation}
For convenience, we define the function
\begin{equation}
f(\phi_k) = \tanh^2\left( \frac{\phi_k}{\sqrt{6\alpha} M_{Pl}} \right),
\end{equation}
which allows for compact expressions of the potential and its derivatives. Its first derivative is given by
\begin{equation}
f'(\phi_k) = \frac{1}{M_{Pl}} \sqrt{\frac{2}{3\alpha}} \sech^2\left( \frac{\phi_k}{\sqrt{6\alpha} M_{Pl}} \right) \tanh\left( \frac{\phi_k}{\sqrt{6\alpha} M_{Pl}} \right).
\end{equation}
These expressions are useful when computing the inflaton mass and the reheating temperature. Using Eqs.~(\ref{mass2}) and (\ref{Trey1}), and setting $\gamma = 1/(3\alpha)$, we obtain
\begin{equation}
\label{massphi}
m_\phi \approx \sqrt{\frac{8\pi^2 A_s}{3\alpha^2}} \, \csch^2\left( \frac{\phi_k}{\sqrt{6\alpha} M_{Pl}} \right) M_{Pl},
\end{equation}
and the reheating temperature (assuming perturbative decay with coupling $y$) becomes
\begin{equation}
\label{Trey2}
T_{re}^{(y)} \approx |y| \left( \frac{15 A_s}{4 \pi^2 g_{re} \alpha^2} \right)^{1/4} \csch\left( \frac{\phi_k}{\sqrt{6\alpha} M_{Pl}} \right) M_{Pl},
\end{equation}
where $g_{re}$ is the effective number of relativistic degrees of freedom at reheating.
Given a set of parameters $(\alpha, y, \omega_{re})$, we numerically solve Eq.~(\ref{Nk1}) for $\phi_k$. This value is then used to compute all relevant inflationary and reheating quantities, such as $n_s$, $r$, $T_{re}$, and $m_\phi$.
We adopt the observational constraints from Planck 2018, BICEP/Keck 2015, and BAO measurements \cite{Akrami:2018odb}
\begin{align}
n_s &= 0.9658 \pm 0.0040 \quad (68\%\, \mathrm{C.L.}), \label{boundsns}\\
r &< 0.068 \quad (95\%\, \mathrm{C.L.}). \label{boundr}
\end{align}
These bounds are used to restrict the allowed parameter space of the model by eliminating the parameter $V_0$ and $\alpha$ in terms of $n_s$ and $r$, as done in \cite{German:2024rmn}.

\begin{figure*}[ht]
\begin{center}
$\begin{array}{ccc}
\includegraphics[width=3.in]{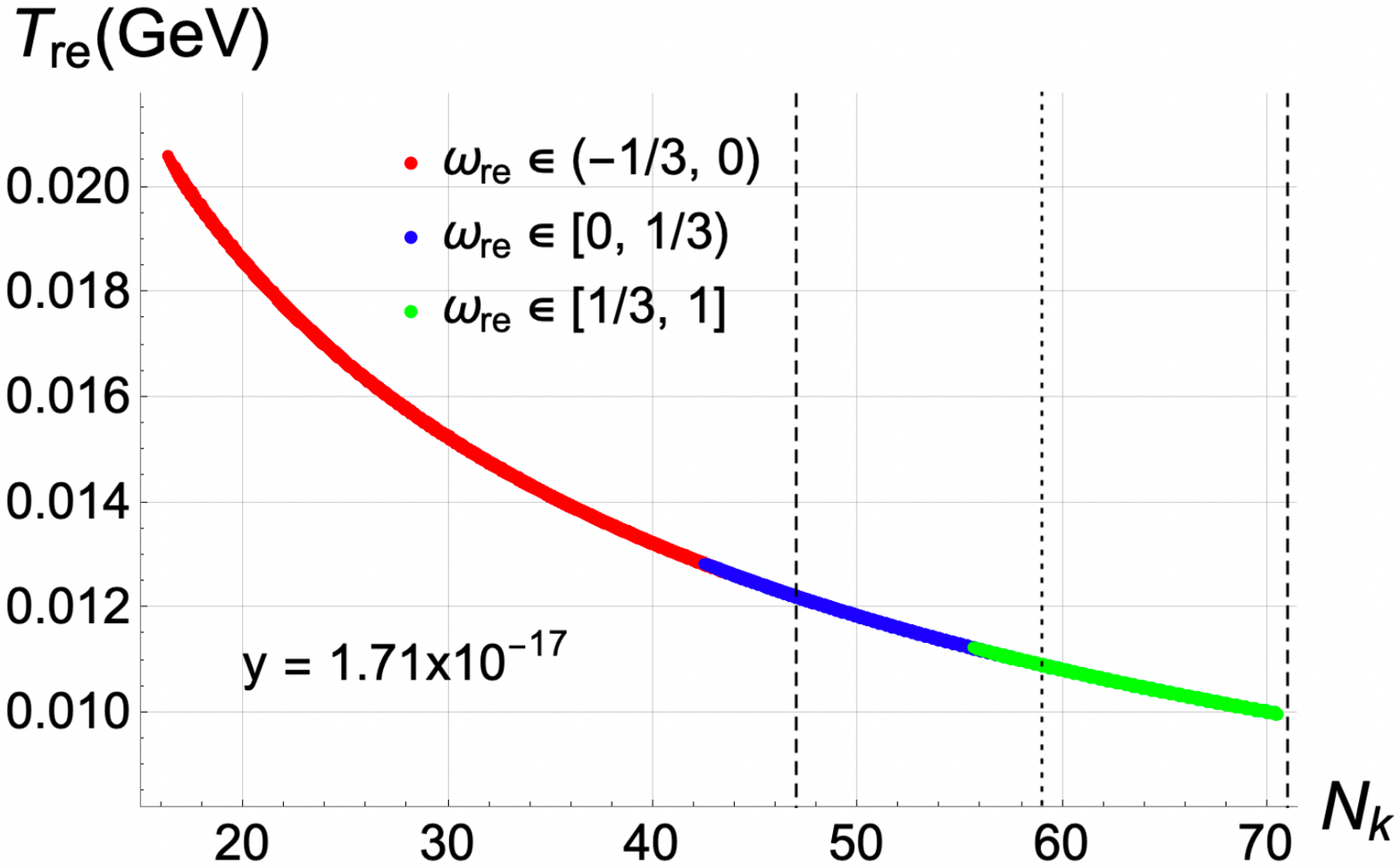}&
\includegraphics[width=3.in]{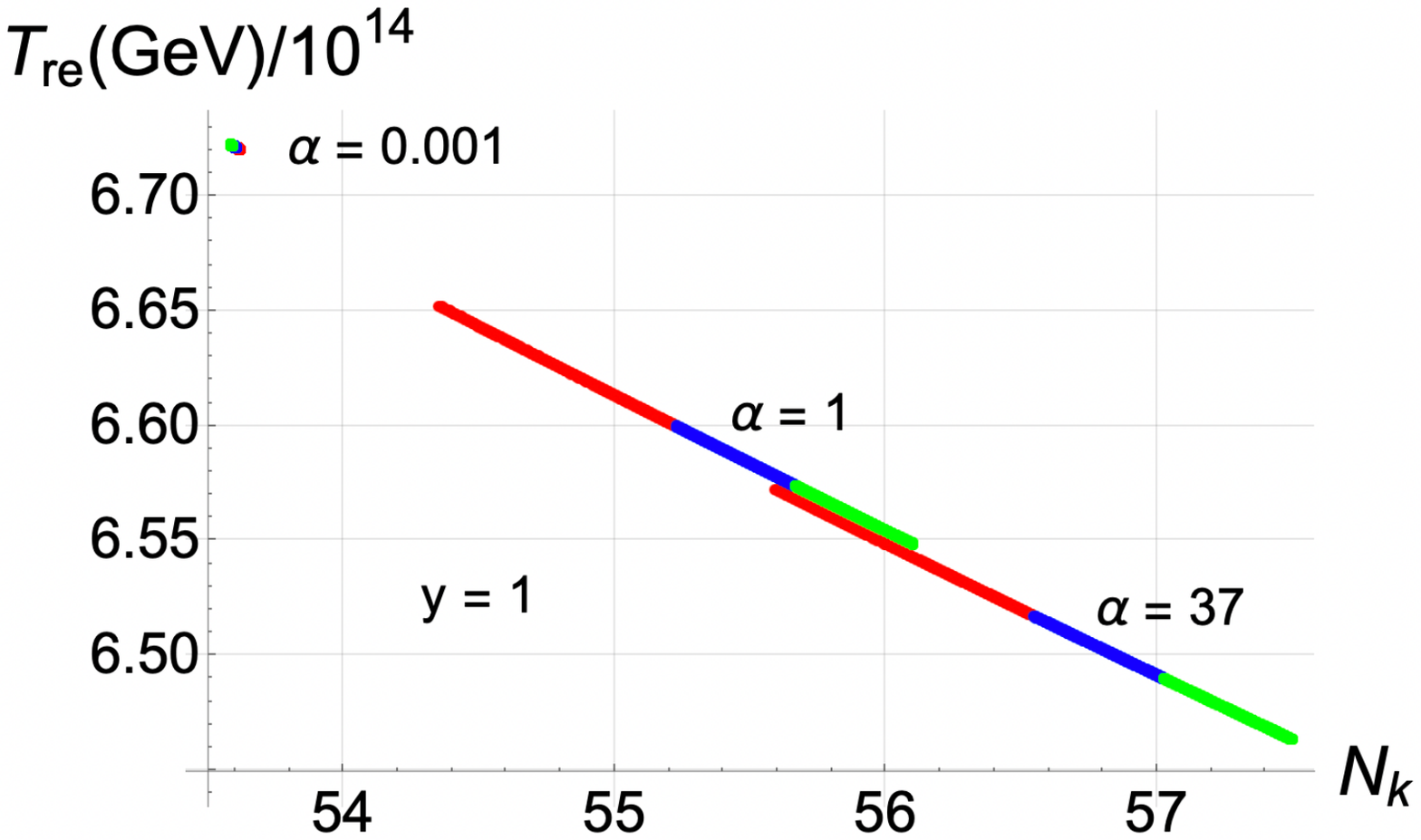}\\
\end{array}$
\caption{The plots show the reheating temperature $T_{re}$, given by Eq.~(\ref{Trey2}), as a function of the number of $e$-folds during inflation, $N_k$, Eq.~(\ref{Nk}), for the $\alpha$-attractor model defined in Eq.~(\ref{pot}). The left panel corresponds to the smallest value of the Yukawa coupling, $y \approx 1.71 \times 10^{-17}$, that is consistent with the lower bound $T_{re} \geq 10~\mathrm{MeV}$. The right panel shows the case $y = 1$, which yields a reheating temperature close to the maximum allowed value, where the number of $e$-folds during reheating is zero. To generate the plots, we first solve Eq.~(\ref{Nk1}) numerically to determine $\phi_k$, using the values shown in the Table~\ref{A_bounds}. All other quantities are then computed from their respective defining equations. The equation-of-state parameter $\omega_{re}$ is varied within the range $-1/3 < \omega_{re} < 1$ and is divided into three intervals, as specified in the left figure. For the small coupling case (left figure, $y \approx 1.71 \times 10^{-17}$), the parameter $\alpha$ is varied continuously in the range $\{0.001, 37\}$. In contrast, for $y = 1$ right figure, we use the discrete values $\alpha = 0.001$, $1$, and $37$, since a continuous range would lead to significant overlap between the curves, making the color coding ineffective. In the small $y$ case, the overlap between curves do not spoil the color coding. The vertical dashed lines indicate the $2\sigma$ confidence interval for $n_s$, obtained from Eq.~(\ref{boundsns}), while the central dotted line corresponds to the mean value of $n_s$, here expressed in terms on $N_k$. Comparing the scales on the axes, we observe that increasing the coupling $y$ leads to a higher reheating temperature $T_{re}$ and to a smaller range of $e$-folds during inflation, $N_k$. A similar trend is observed in Figs.~\ref{A_Nk_ns_y} to \ref{A_r_ns_y}, where the quantities are plotted as functions of the scalar spectral index $n_s$ instead of $N_k$. These figures correspond to the same extreme values of $y$ used in Fig.~\ref{A_Tre_Nk_y}, and are included to highlight features that are not visible when plotting $\log_{10}(T_{re})$ directly, as is clearly illustrated in the Fig.~\ref{A_LogTre_Nk_y} below.
}
\label{A_Tre_Nk_y}
\end{center}
\end{figure*}

\begin{figure*}[ht]
\begin{center}
\includegraphics[width=4.5 in]{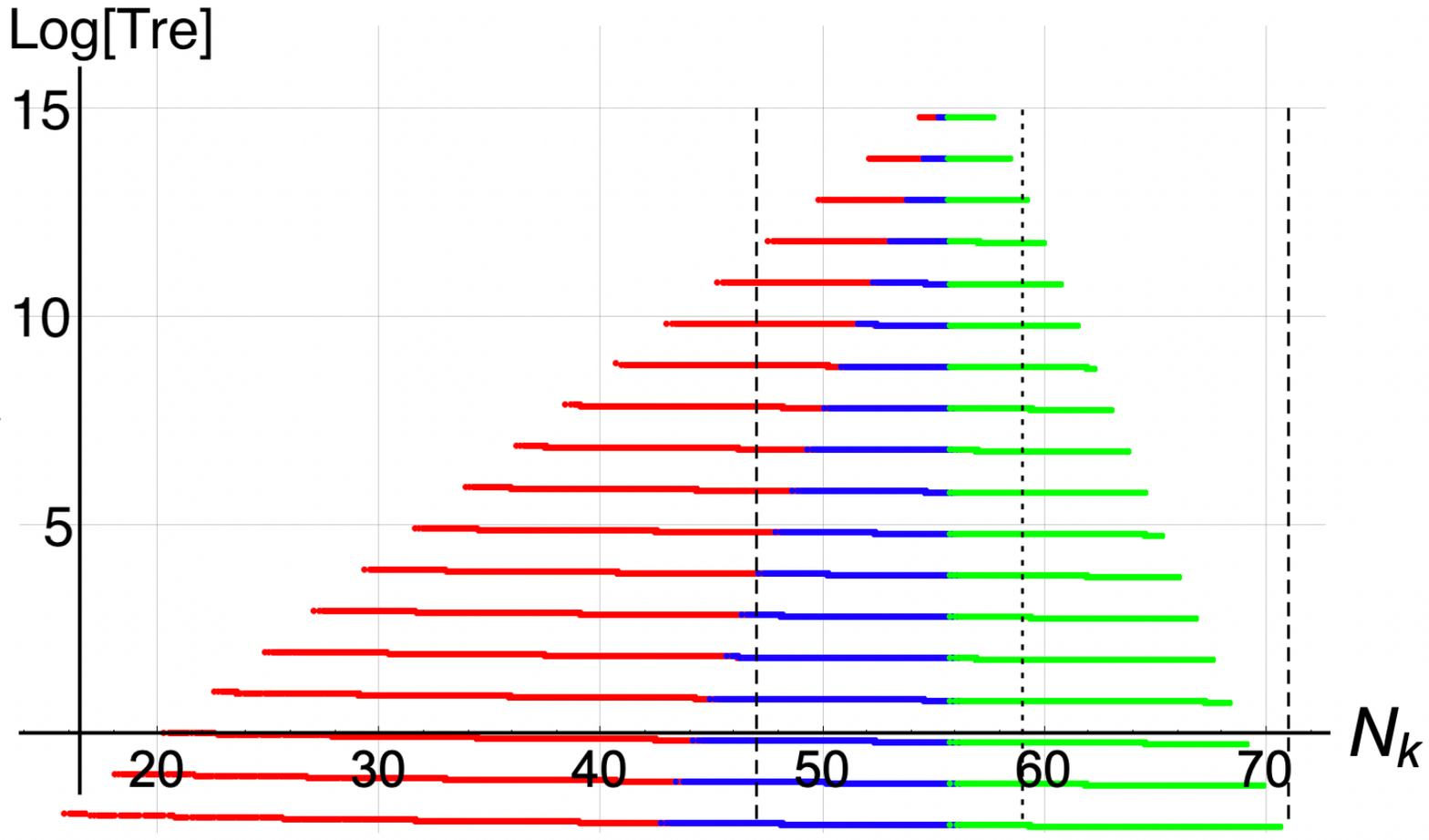}
\caption{The figure shows $\log_{10}(T_{re})$ for values of the Yukawa coupling ranging from $y \approx 1.71 \times 10^{-17}$ (lower curve) to $y = 1$ (upper curve). As the coupling $y$ increases, the reheating temperature $T_{re}$, given by Eq.~(\ref{Trey2}), becomes larger, while the range for the number of $e$-folds during inflation $N_k$, Eq.~(\ref{Nk}), decreases. All other details and parameter choices are the same as in Fig.~\ref{A_Tre_Nk_y}.}
\label{A_LogTre_Nk_y}
\end{center}
\end{figure*}

\begin{figure*}[ht]
\begin{center}
$\begin{array}{ccc}
\includegraphics[width=3.in]{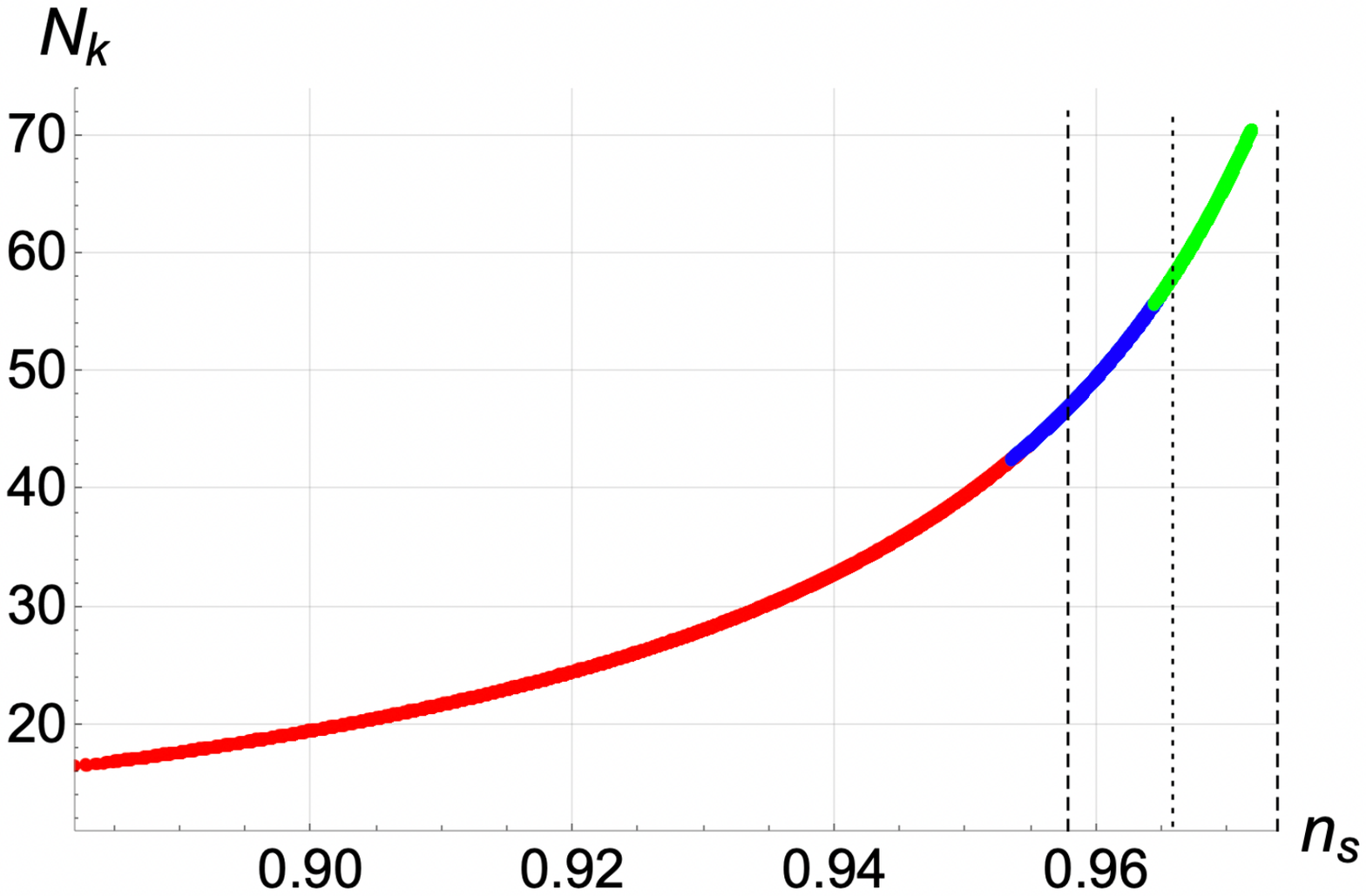}&
\includegraphics[width=3.in]{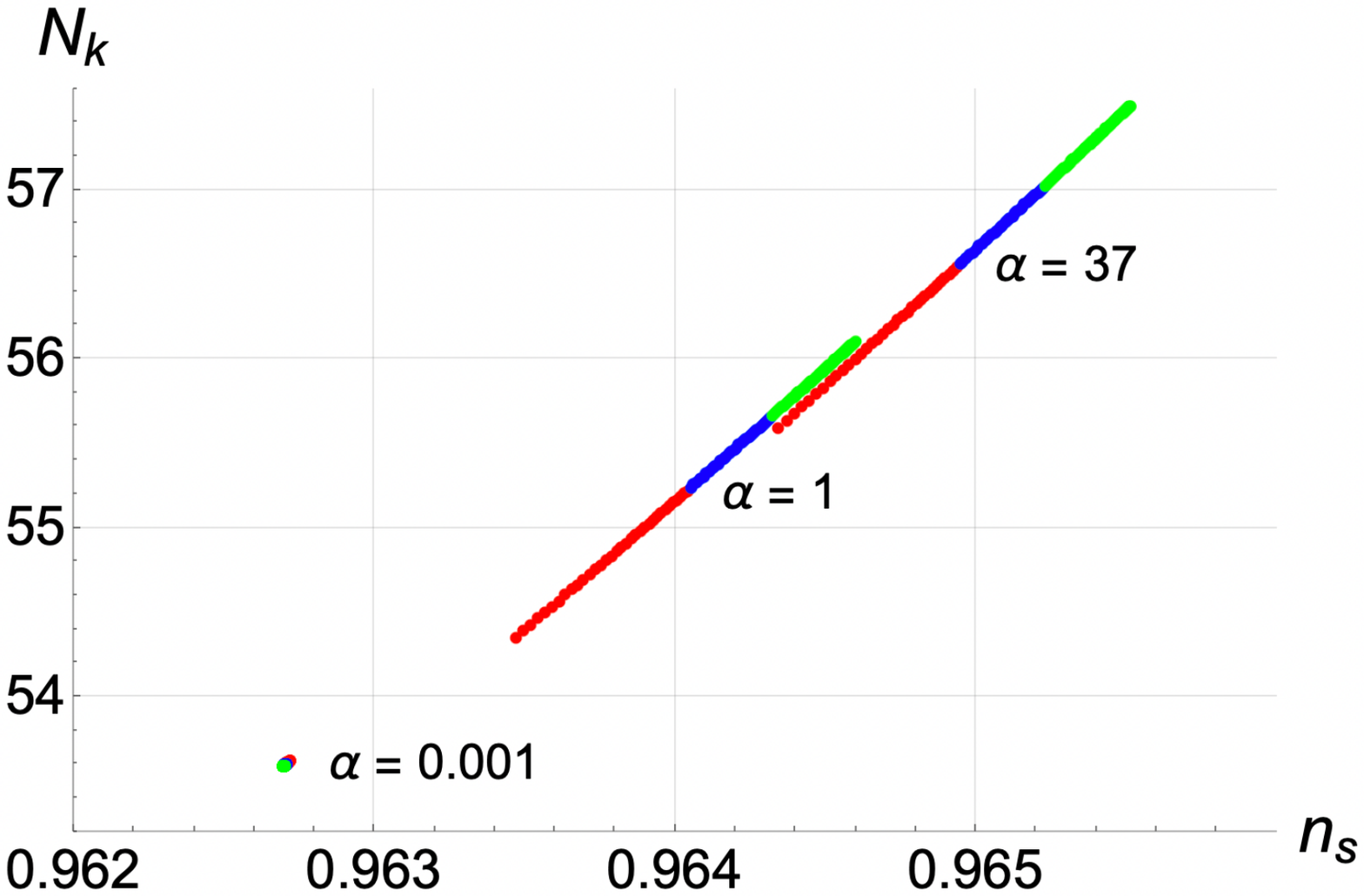}\\
\end{array}$
\caption{The plots show the number of $e$-folds during inflation $N_k$, given by Eq.~(\ref{Nk}), as a function of the scalar spectral index $n_s$, which follows from Eq.~(\ref{ns}). All other details are the same as in Fig.~\ref{A_Tre_Nk_y}.}
\label{A_Nk_ns_y}
\end{center}
\end{figure*}

\begin{figure*}[ht]
\begin{center}
$\begin{array}{ccc}
\includegraphics[width=3.in]{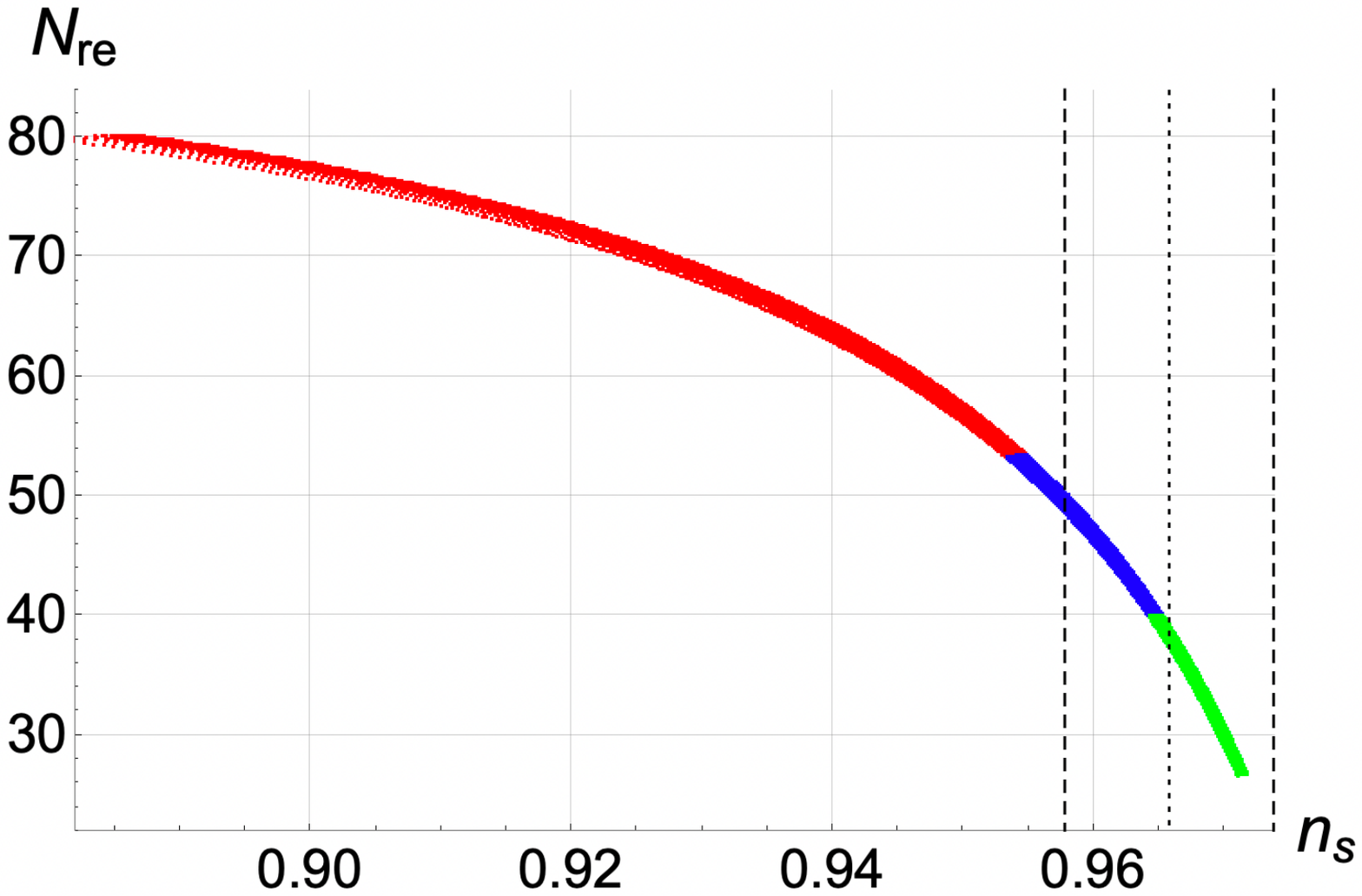}&
\includegraphics[width=3.in]{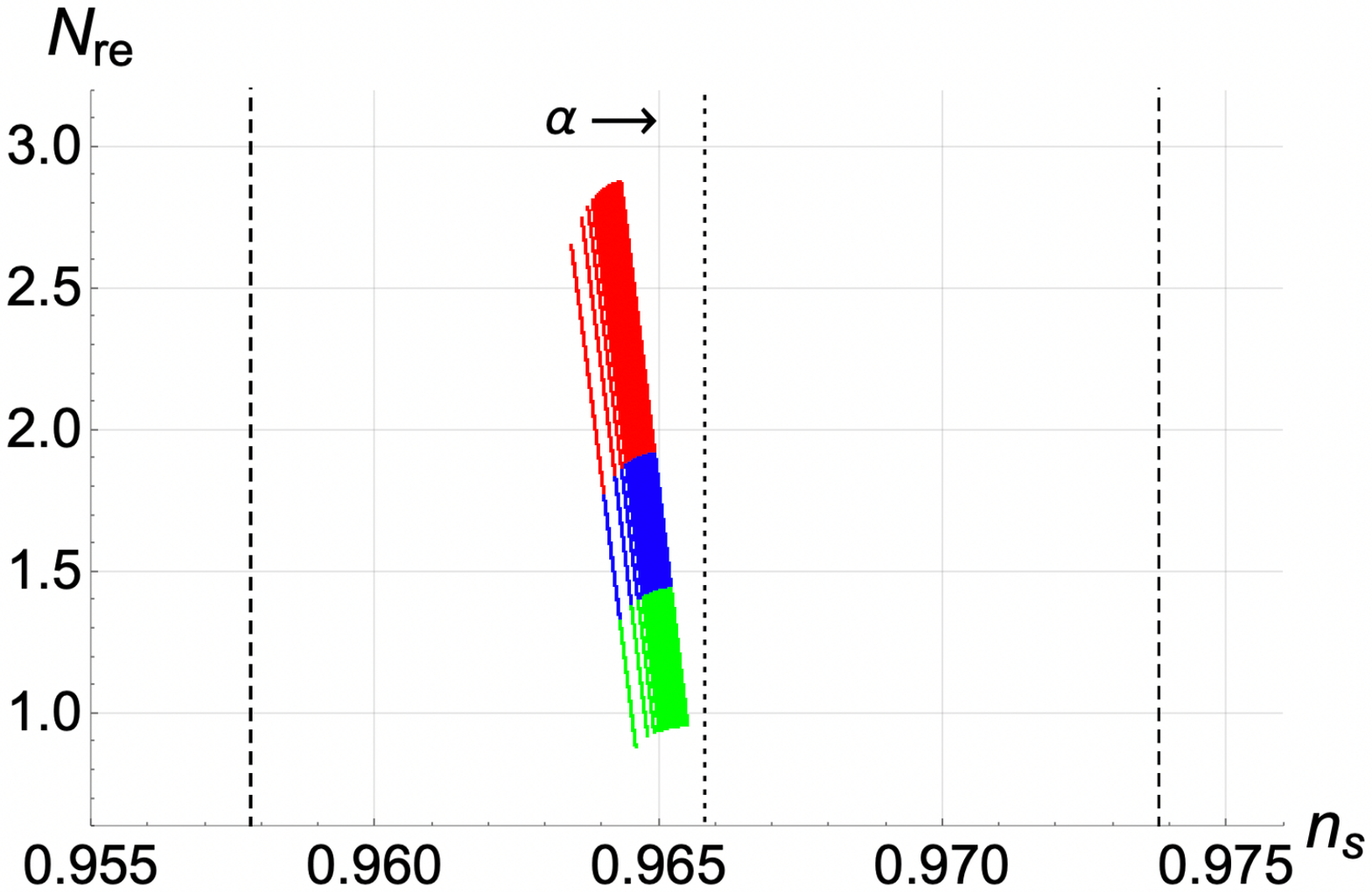}\\
\end{array}$
\caption{The plots show the number of $e$-folds during reheating $N_{re}$, which follows from Eq.~(\ref{Nre}), as a function of the scalar spectral index $n_s$, given by Eq.~(\ref{ns}). All other details are the same as in Fig.~\ref{A_Tre_Nk_y}.}
\label{A_Nre_ns_y}
\end{center}
\end{figure*}

\begin{figure*}[ht]
\begin{center}
$\begin{array}{ccc}
\includegraphics[width=3.in]{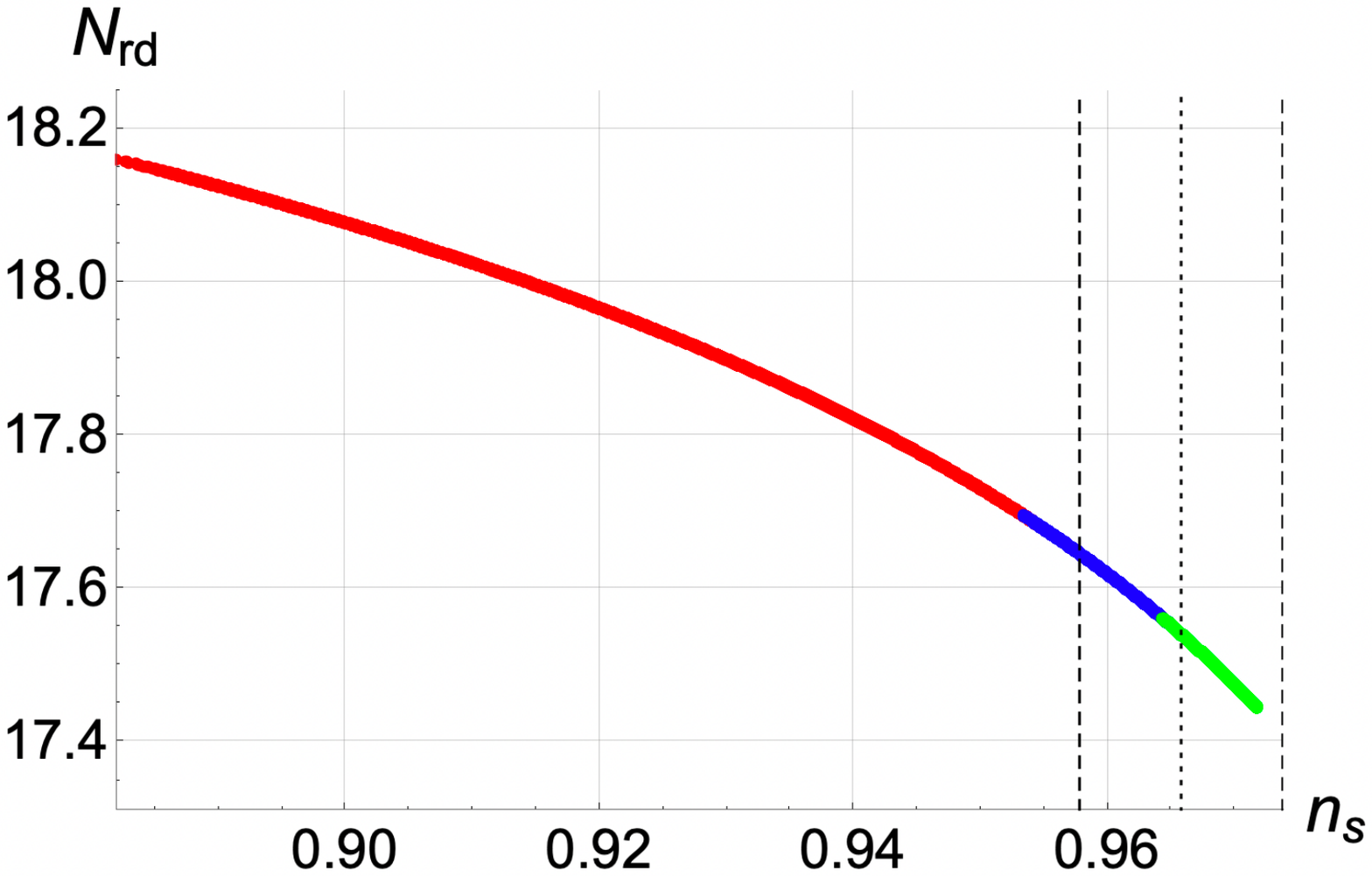}&
\includegraphics[width=3.in]{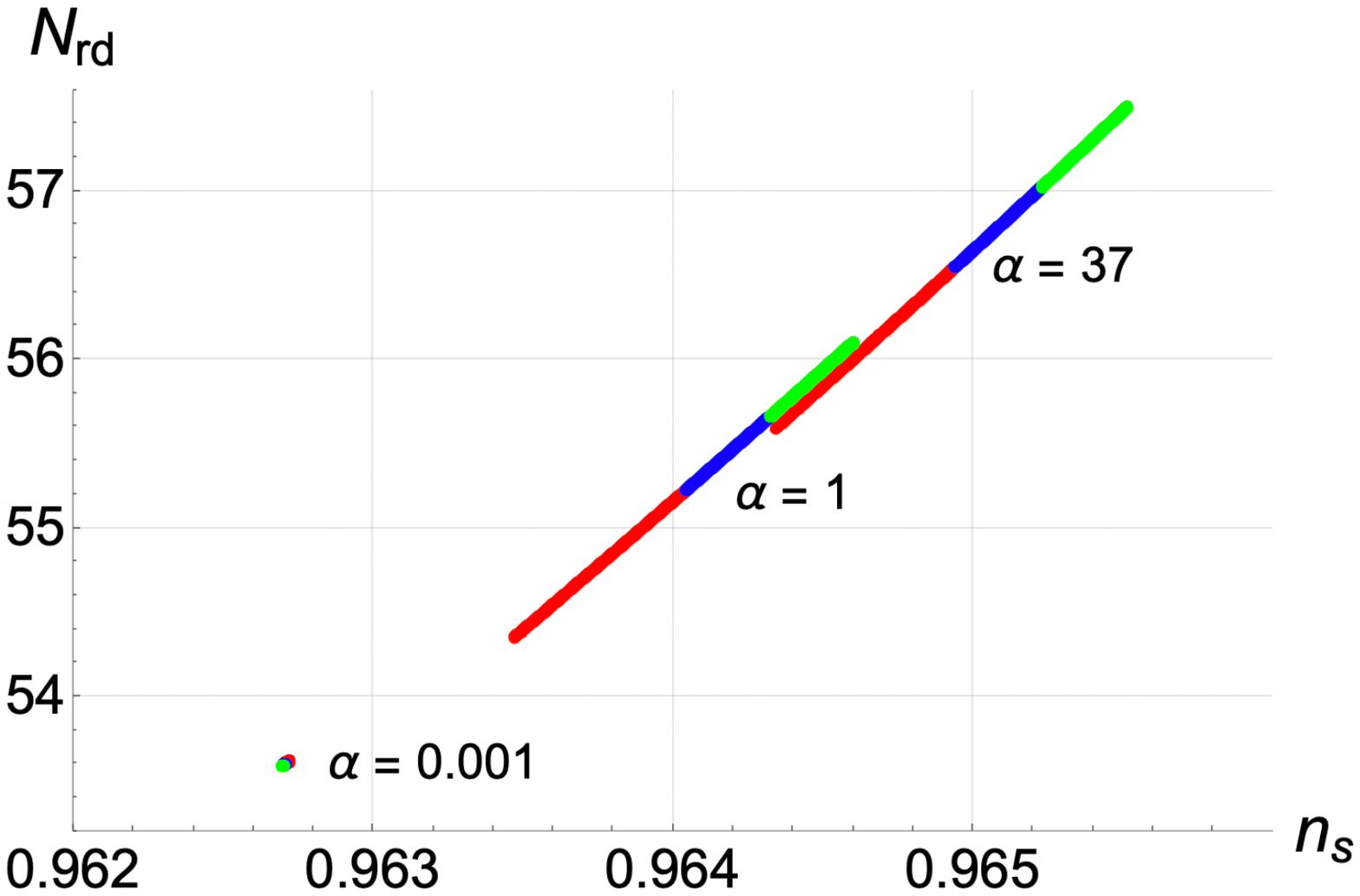}\\
\end{array}$
\caption{The plots show the number of $e$-folds during the radiation dominated epoch $N_{rd}$, given by $N_{rd} \equiv \ln\left(\frac{a_{eq}}{a_r}\right) = \ln\left(\frac{a_{eq} T_{re}}{\left(\frac{43}{11 g_{s,re}}\right)^{1/3} a_0 T_0}\right)$, as a function of the scalar spectral index $n_s$, given by Eq.~(\ref{ns}). All other details are the same as in Fig.~\ref{A_Tre_Nk_y}.}
\label{A_Nrd_ns_y}
\end{center}
\end{figure*}

\begin{figure*}[ht]
\begin{center}
$\begin{array}{ccc}
\includegraphics[width=3.in]{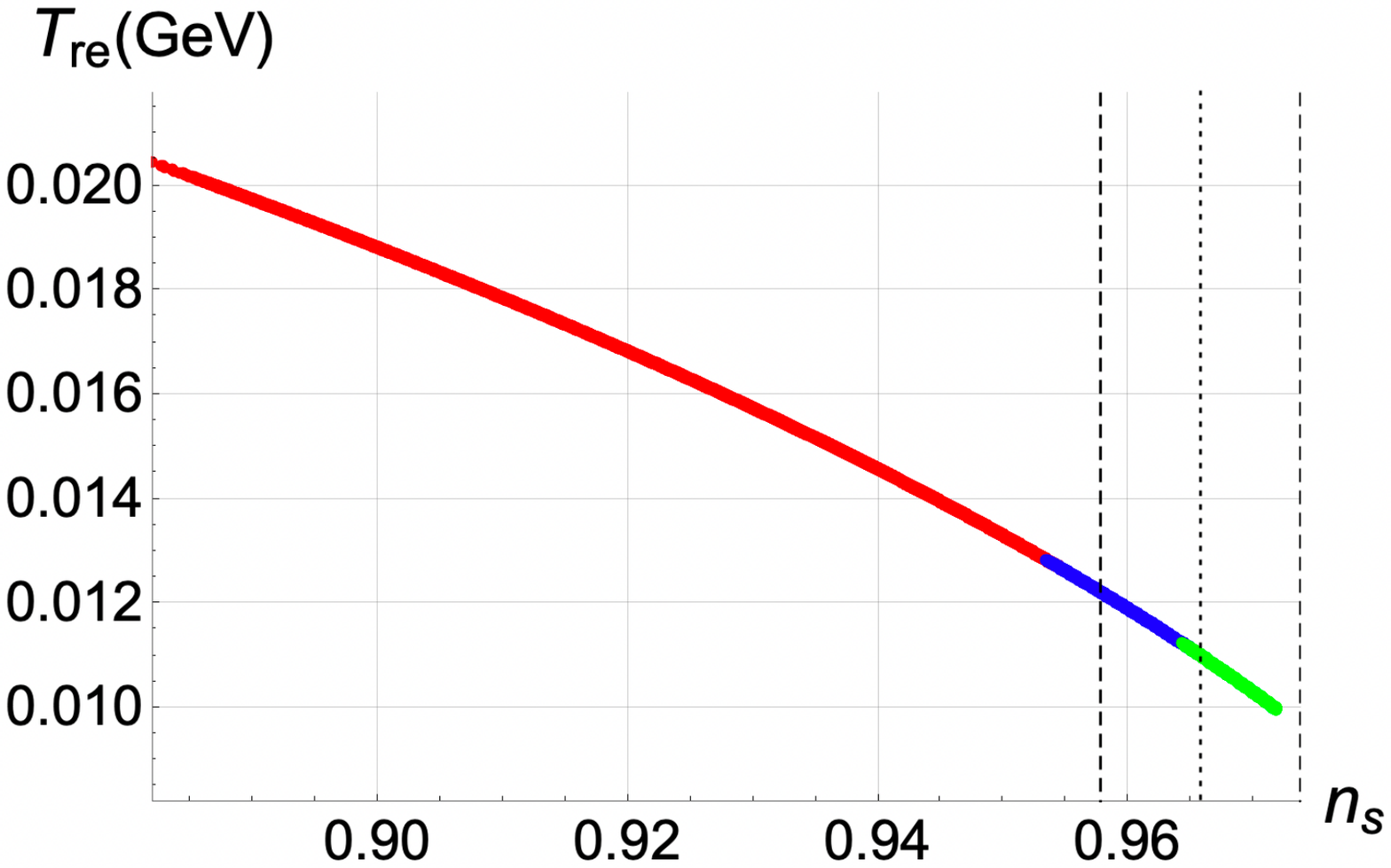}&
\includegraphics[width=3.in]{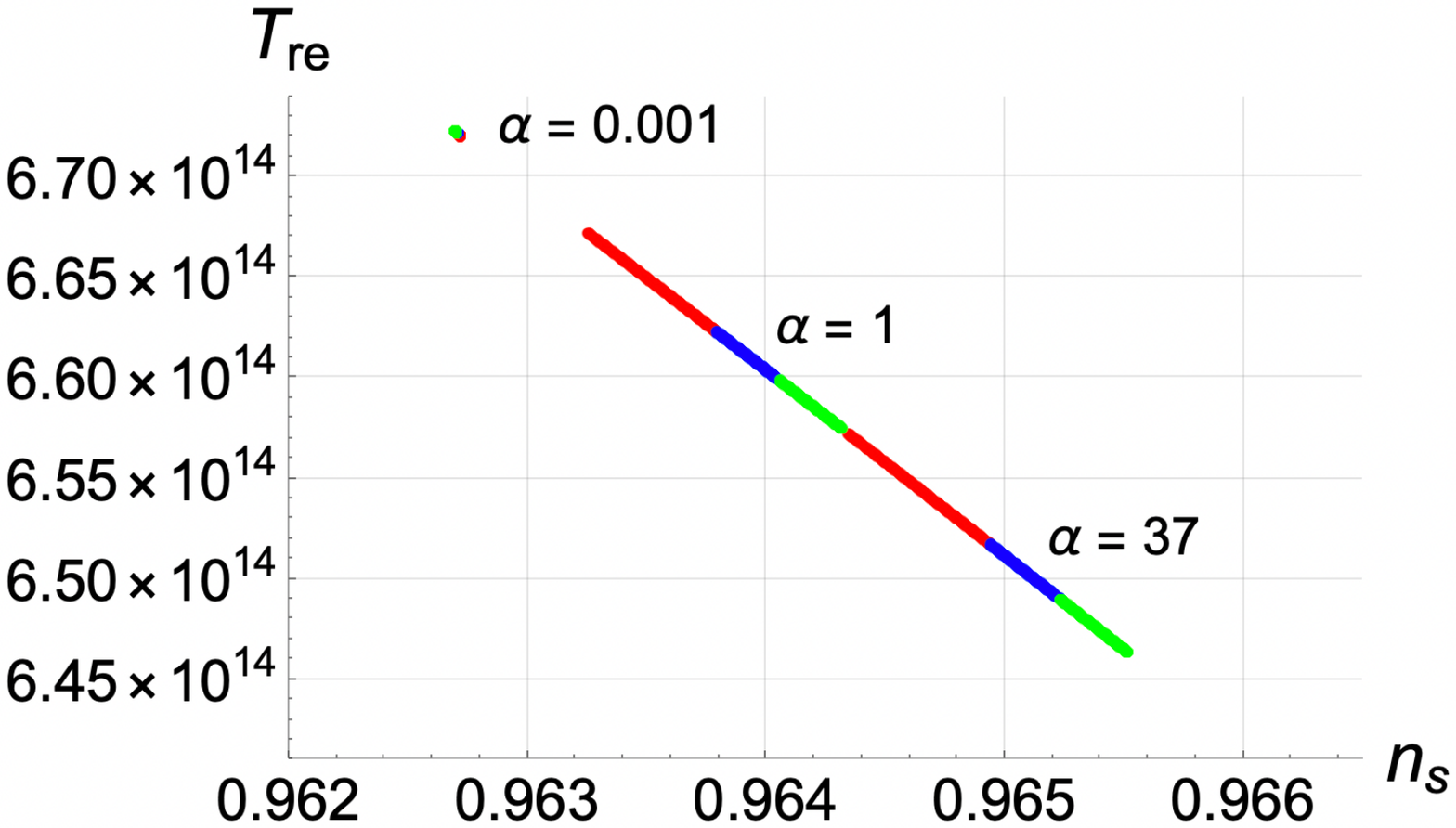}\\
\end{array}$
\caption{The plots show the reheating temperature $T_{re}$, given by Eq.~(\ref{Trey2}), as a function of the scalar spectral index $n_s$, Eq.~(\ref{ns}). All other details are the same as in Fig.~\ref{A_Tre_Nk_y}.}
\label{A_Tre_ns_y}
\end{center}
\end{figure*}

The main results of this work are presented in the figures Fig.~\ref{A_Tre_Nk_y} to Fig.~\ref{A_nsk_y}. Figure~\ref{A_Tre_Nk_y} shows the reheating temperature $T_{re}$ for two representative values of the Yukawa coupling: the lower bound $y \approx 1.71 \times 10^{-17}$, obtained from the condition $T_{re} > 10\,\mathrm{MeV}$,
\footnote{The lower bound $T_{re} > 10\,\mathrm{MeV}$ is motivated by constraints from Big Bang Nucleosynthesis (BBN). BBN occurs at temperatures of a few MeV, and its successful predictions of light element abundances require that the universe be radiation-dominated at that time. If reheating were completed at temperatures significantly below $\sim 10\,\mathrm{MeV}$, the resulting thermal history could alter neutron-to-proton ratios and nuclear reaction rates, leading to discrepancies with observed abundances of light nuclei. The bound $T_{re} > 10\,\mathrm{MeV}$ is therefore a conservative threshold that ensures consistency with BBN. In this work, it is used for definiteness; adopting a slightly lower value (e.g., $T_{re} > 4\,\mathrm{MeV}$) would not significantly affect the results, as the reheating temperatures predicted by the models considered lie well above this limit.} and the reference value $y = 1$. Figure~\ref{A_LogTre_Nk_y} displays $\log_{10}(T_{re}/\mathrm{GeV})$ over the full range of $y$ and $\omega_{re}$ values, following the same conventions as in Fig.~\ref{A_Tre_Nk_y}. Figures~\ref{A_Nk_ns_y} to~\ref{A_r_ns_y} show the evolution of the number of $e$-folds during inflation $N_k$, reheating $N_{re}$, and radiation $N_{rd}$, as well as the reheating temperature $T_{re}$, the tensor-to-scalar ratio $r$, and the running of the spectral index $n_{sk}$, all as functions of the scalar spectral index $n_s$. These are shown for the two extreme values of $y$, and for selected values of $\omega_{re}$. Finally, Figs.~\ref{A_Nk_y} to~\ref{A_nsk_y} include, in addition to the previously mentioned cosmological quantities, the inflaton mass as a function of $y$. As stated in the introduction, the goal of this analysis is to study the general behavior and parametric dependence of these quantities. We do not focus on the precision of the approximations used, as they do not significantly affect the qualitative trends observed. Moreover, no statistical data analysis is performed. This work should be regarded as a complementary study to analyses such as the one presented in~\cite{Ellis:2021kad}, where also gravitino overproduction constraints after inflation are considered.

In supergravity theories, the gravitino, the spin-$\frac{3}{2}$ superpartner of the graviton, plays an important role in the post-inflationary cosmological evolution. Depending on its mass and lifetime, the gravitino can have significant implications for Big Bang Nucleosynthesis (BBN), the cosmic microwave background (CMB), and the dark matter abundance \cite{Khlopov:1984pf,Ellis:1984eq,Kawasaki:2004qu}. Gravitinos can be produced thermally during the reheating phase, via scatterings of particles in the thermal plasma. The resulting number density is approximately proportional to the reheating temperature $T_{re}$ \cite{Bolz:2000fu}, \cite{Pradler:2006qh}
\begin{equation}
Y_{3/2} \equiv \frac{n_{3/2}}{s} \propto \frac{T_{re}}{M_{Pl}},
\end{equation}
where $n_{3/2}$ is the gravitino number density and $s$ is the entropy density. As a result, high reheating temperatures can lead to an overabundance of gravitinos. If the gravitino is unstable and decays after BBN, its decay products (including photons and hadrons) can disrupt the formation of light elements, leading to strong constraints on its abundance and therefore on $T_{\mathrm{re}}$ \cite{Kawasaki:2004qu}, \cite{Cyburt:2009pg}. For gravitino masses in the range $m_{3/2} \sim 100~\mathrm{GeV} - 10~\mathrm{TeV}$, the corresponding upper bound on the reheating temperature is approximately
\begin{equation}
T_{re} \lesssim 10^6 - 10^9~\mathrm{GeV}.
\end{equation}
If the gravitino is stable (e.g., in models where it is the lightest supersymmetric particle, LSP), its relic abundance must not exceed the observed dark matter density. This also imposes an upper limit on $T_{re}$ \cite{Moroi:1993mb}, \cite{Bolz:2000fu}. Therefore, in supersymmetric models with supergravity, the requirement of avoiding gravitino overproduction provides an important upper bound on $T_{re}$. In our work, imposing constraints on $T_{re}$ would be reflected in constraints on the Yukawa coupling $y$ but this has already been done in~\cite{Ellis:2021kad}.
\section{Non-Perturbative Reheating Effects in \boldmath{$\alpha$}-Attractor Models}\label{preheating}
In addition to the perturbative decay processes discussed in the previous sections, reheating may also proceed through non-perturbative mechanisms, collectively known as \textit{preheating}  \cite{Kofman:1994rk}, \cite{Felder:2001kt} \cite{Micha:2002ey}. These mechanisms rely on the coherent oscillations of the inflaton field around the minimum of its potential and can lead to explosive particle production via processes such as parametric resonance or tachyonic instability. In the context of single-field inflation, preheating can occur when the inflaton couples to other scalar or fermionic fields through interaction terms like $g^2 \phi^2 \chi^2$ or $y \phi \bar{\psi} \psi$. Unlike perturbative reheating, where the inflaton decays gradually, non-perturbative effects can transfer a significant fraction of the inflaton's energy into other fields within just a few oscillations.

For bosonic couplings of the form
\begin{equation}
\mathcal{L}_{\text{int}} = -\frac{1}{2} g^2 \phi^2 \chi^2,
\end{equation}
the time-dependent inflaton oscillations induce a periodically varying effective mass for the $\chi$ field
\begin{equation}
m_\chi^2(t) = g^2 \phi^2(t),
\end{equation}
which can lead to broad parametric resonance depending on the coupling strength $g$ and the amplitude of the inflaton oscillations. In this regime, certain modes of the $\chi$ field grow exponentially, resulting in an efficient and rapid transfer of energy from the inflaton to $\chi$ particles. Alternatively, in models with spontaneous symmetry breaking, such as hybrid inflation, the effective mass of decay products may become negative during certain phases of the inflaton oscillations. This leads to tachyonic preheating \cite{Felder:2000hj}, characterized by the rapid growth of long-wavelength modes due to instabilities in the potential.

In $\alpha$-attractor models, the potential flattens at large field values, a feature that enhances agreement with current CMB constraints. After inflation ends, the inflaton field rolls toward the minimum of the potential near $\phi = 0$ and undergoes damped oscillations. The shape of the potential near the minimum is approximately quadratic
\begin{equation}
V(\phi) \approx \frac{1}{2} m_\phi^2 \phi^2,
\end{equation}
with $m_\phi$ determined by the curvature of the potential at the minimum. This behavior supports the possibility of preheating through parametric resonance for sufficiently strong couplings. However, the efficiency of preheating in $\alpha$-attractors depends on the value of $\alpha$. For small $\alpha$, the potential near the minimum is steeper, resulting in higher oscillation frequencies and larger inflaton amplitudes, which can enhance resonance. In contrast, for large $\alpha$, the potential becomes flatter, and the inflaton oscillations are slower and less energetic, possibly suppressing non-perturbative effects. In addition, the suppression of higher-order interactions in supergravity-inspired $\alpha$-attractors may limit the available decay channels, favoring perturbative reheating unless additional couplings are introduced. Therefore, both the geometric structure of the scalar manifold (encoded in $\alpha$) and the UV completion of the model play a crucial role in determining the importance of preheating.
\section{Discussion and Conclusions}\label{Con}
In this work, we developed a general analytical framework to study the reheating phase following inflation in single-field models with perturbative inflaton decay. We obtained expressions for the reheating temperature $T_{re}$, the number of $e$-folds during reheating $N_{re}$, and the inflaton mass $m_\phi$ among other cosmological quantities of interest for three representative types of couplings: gravitational, scalar, and Yukawa-type fermionic interactions. We applied this framework to $\alpha$-attractor inflationary models,  characterized by a potential of the form $V(\phi) = V_0 \tanh^2(\phi/\sqrt{6\alpha} M_{Pl})$. These models exhibit attractor behavior and are compatible with current bounds on the scalar spectral index $n_s$ and the tensor-to-scalar ratio $r$. Focusing on fermionic decay, we derived an explicit expression for $T_{re}$ in terms of the Yukawa coupling $y$ and the $\alpha$ parameter. By solving the reheating condition  Eq.~(\ref{Nk1}) numerically, we determined the inflaton field value at horizon exit $\phi_k$, and computed all associated cosmological quantities. We then examined how these quantities behave as functions of the scalar spectral index $n_s$ and $y$, for a range of values of $\alpha$, $y$, and $\omega_{re}$. The emphasis of this work is on understanding the qualitative evolution of the reheting temperature $T_{re}$,  inflaton mass $m_\phi$, the number of $e$-folds during inflation $N_{k}$, reheating $N_{re}$, and radiation $N_{rd}$, as well as the scalar spectral index $n_s$, its running $n_{sk}$, and the tensor-to-scalar ratio $r$ with respect to $n_s$ and $y$. Although the analytical formulas employed are approximate, they faithfully capture the relevant trends and dependencies. The numerical results and figures illustrate how these quantities vary across observationally allowed ranges of $n_s$ and coupling strengths. Several interesting features can be read directly from the figures, which complement work on data analysis and precise numerical studies of specific topics. This approach can be extended to other inflationary potentials or decay channels, and provides a way to explore how cosmological observables are influenced by the underlying parameters governing the dynamics of the early universe. Non-perturbative reheating mechanisms, such as preheating, can also be incorporated in future work to further refine the picture.

Finally, a brief consideration of the recent results of the ACT experiment is required. When combined with Planck and BAO data, the Atacama Cosmology Telescope (ACT) reported in its 2025 Data Release 6 (DR6) a scalar spectral index of $n_s = 0.9743 \pm 0.0034$, higher than Planck's 2018 value of $n_s = 0.9649 \pm 0.0042$ (without running) \cite{ACT:2025fju}, \cite{ACT:2025tim}.  The central values differ by  $\Delta n_s \approx 0.0094$, approximately a deviation of $\sim 2\sigma$ depending on the model and data set used. This discrepancy may arise from several sources: instrumental systematics, given that ACT is ground-based and Planck was space-based; differences in sky coverage, with Planck observing the full sky and ACT a limited region; variations in statistical methods, such as prior choices and model assumptions e.g., lensing, neutrinos, $N_{\rm eff}$; or new physics beyond $\Lambda$CDM. Although moderate, the tension could become significant if supported by further high-precision measurements. It may then suggest limitations in current models or indicate missing systematics. Reliable parameter estimation from CMB data depends on accurate calibration, beam modeling, foreground removal, and noise characterization. Errors in any of these components can bias the results, such as $n_s$. Further analysis with combined data sets and better control of systematics are needed to clarify the origin of the discrepancy. From our figures plotted against $n_s$, it is clear that the region favored by ACT is marginally reached for very small $y$ couplings and $\omega_{re}$ close to one (green color). If the ACT results for $n_s$ are confirmed, the model analyzed here will probably be discarded.
\acknowledgments

We would like to thank DGAPA-PAPIIT grant: IN110325 {\it Estudios en cosmolog\'ia inflacionaria, agujeros negros primordiales y
energ\'ia oscura}, UNAM, and the SECIHTI ``SNII''  for its funding and support. 

\newpage


\begin{table*}[ht!]
 \begin{center}
{\begin{tabular}{cccc}
\small
Parameter & Value & Parameter & Value  \\
\hline \hline
{$g_{re} $} & $106.75$ & {$k_p $} & $0.05Mpc^{-1}$ \\
{$g_{s,re} $} & $106.75$ & {$T_0 $}   & $2.7255\,K$  \\
{$A_s $}  & $2.1\times 10^{-9}$ & {$a_{eq} $}& $2.94\times 10^{-4}$ \\
 \hline \hline
\end{tabular}}
\caption{The table above provides the parameter values used in the calculations. 
}
\label{A_bounds}
\end{center}
\end{table*}
\begin{figure*}[ht]
\begin{center}
$\begin{array}{ccc}
\includegraphics[width=3.in]{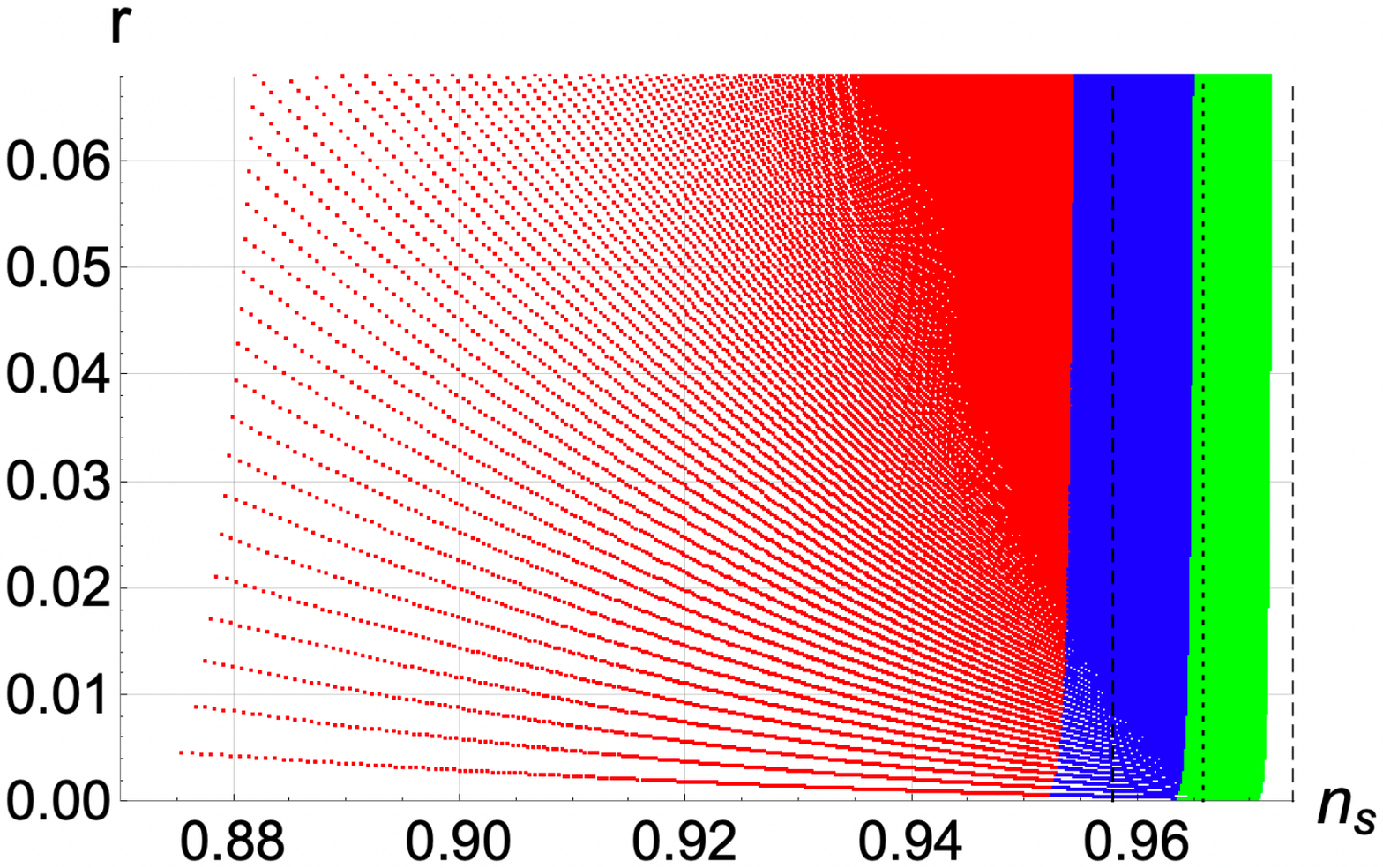}&
\includegraphics[width=3.in]{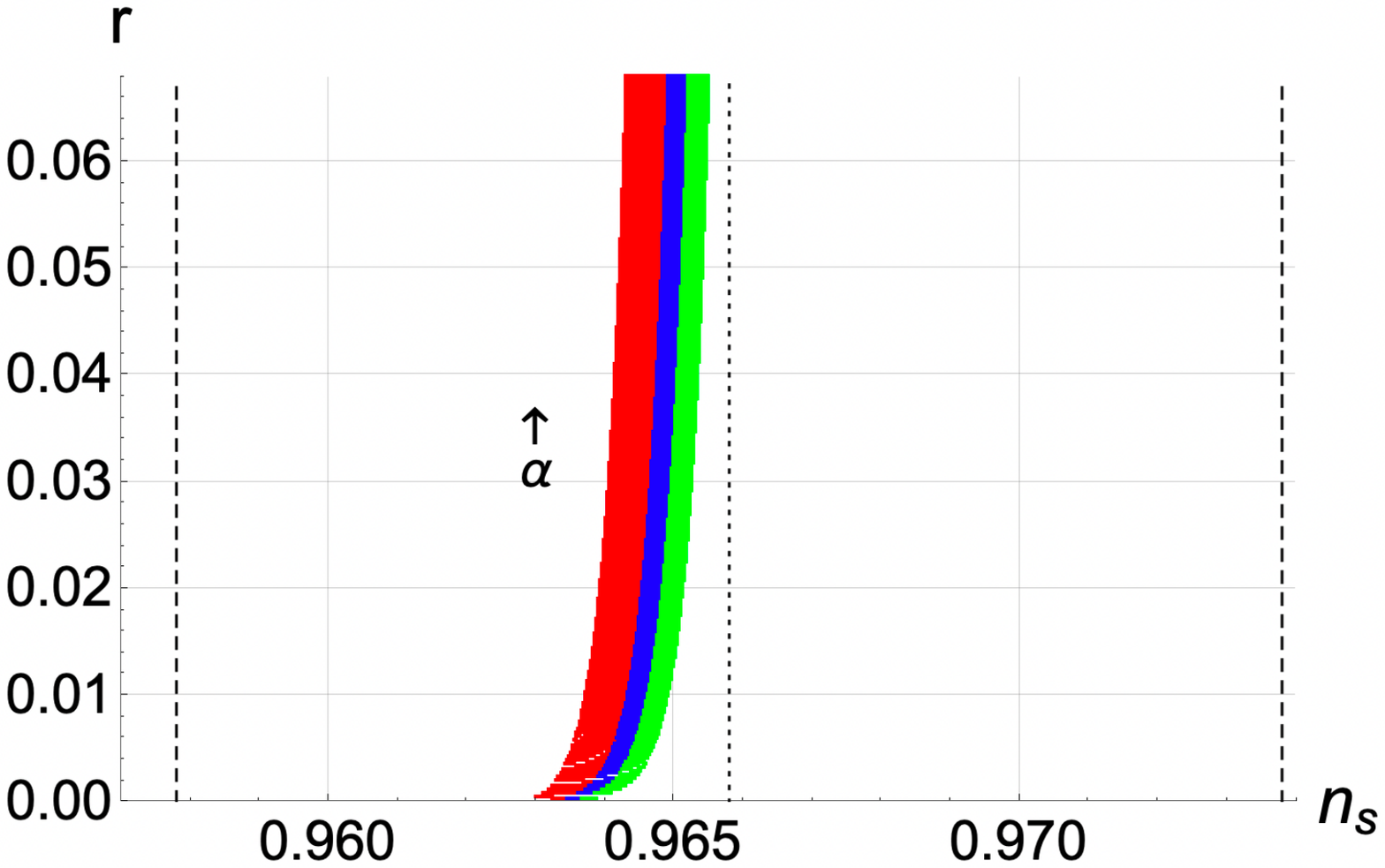}\\
\end{array}$
\caption{The plots show the tensor-to-scalar ratio $r$, given by Eq.~(\ref{r}),  as a function of the scalar spectral index $n_s$, Eq.~(\ref{ns}). All other details are the same as in Fig.~\ref{A_Tre_Nk_y}.}
\label{A_r_ns_y}
\end{center}
\end{figure*}
\begin{figure*}[ht]
\begin{center}
$\begin{array}{ccc}
\includegraphics[width=3.in]{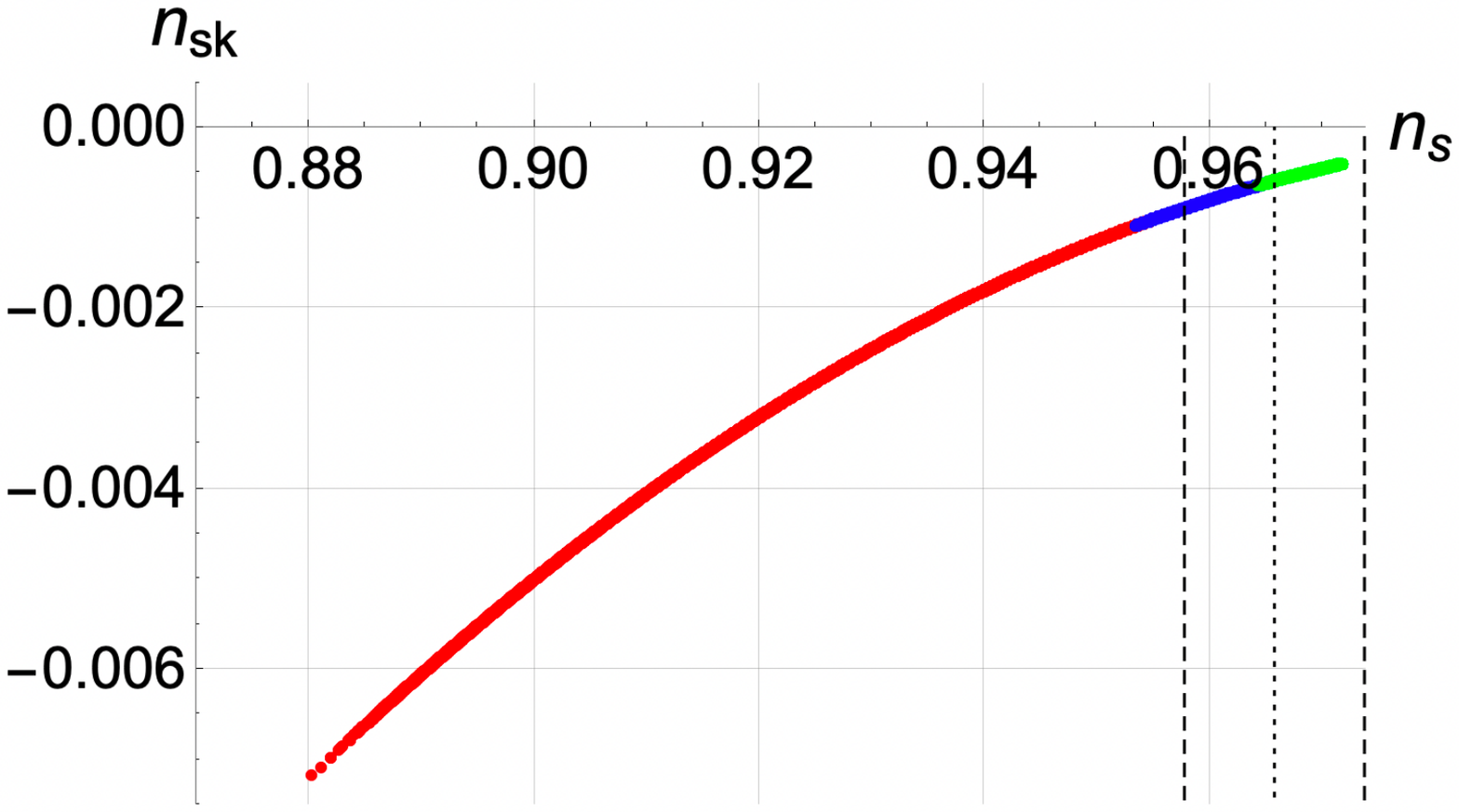}&
\includegraphics[width=3.in]{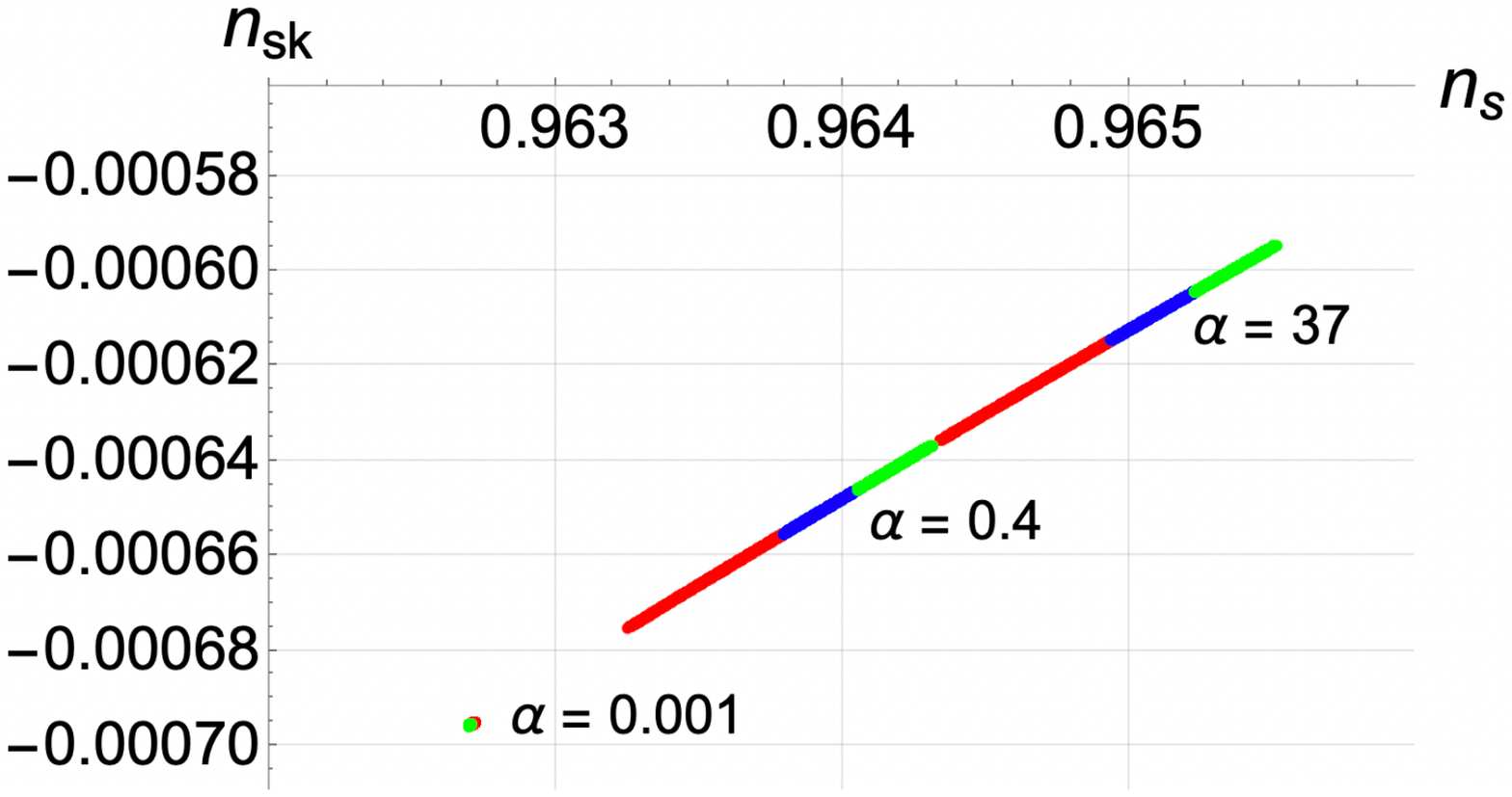}\\
\end{array}$
\caption{The plots show the running of the scalar spectral index, $n_{sk}$ (commonly denoted by $\alpha$), given by Eq.~(\ref{nsk}), as a function of the scalar spectral index $n_s$, Eq.~(\ref{ns}). All other details are the same as in Fig.~\ref{A_Tre_Nk_y}.}
\label{A_r_ns_y}
\end{center}
\end{figure*}
\begin{figure*}[ht]
\begin{center}
$\begin{array}{ccc}
\includegraphics[width=3.in]{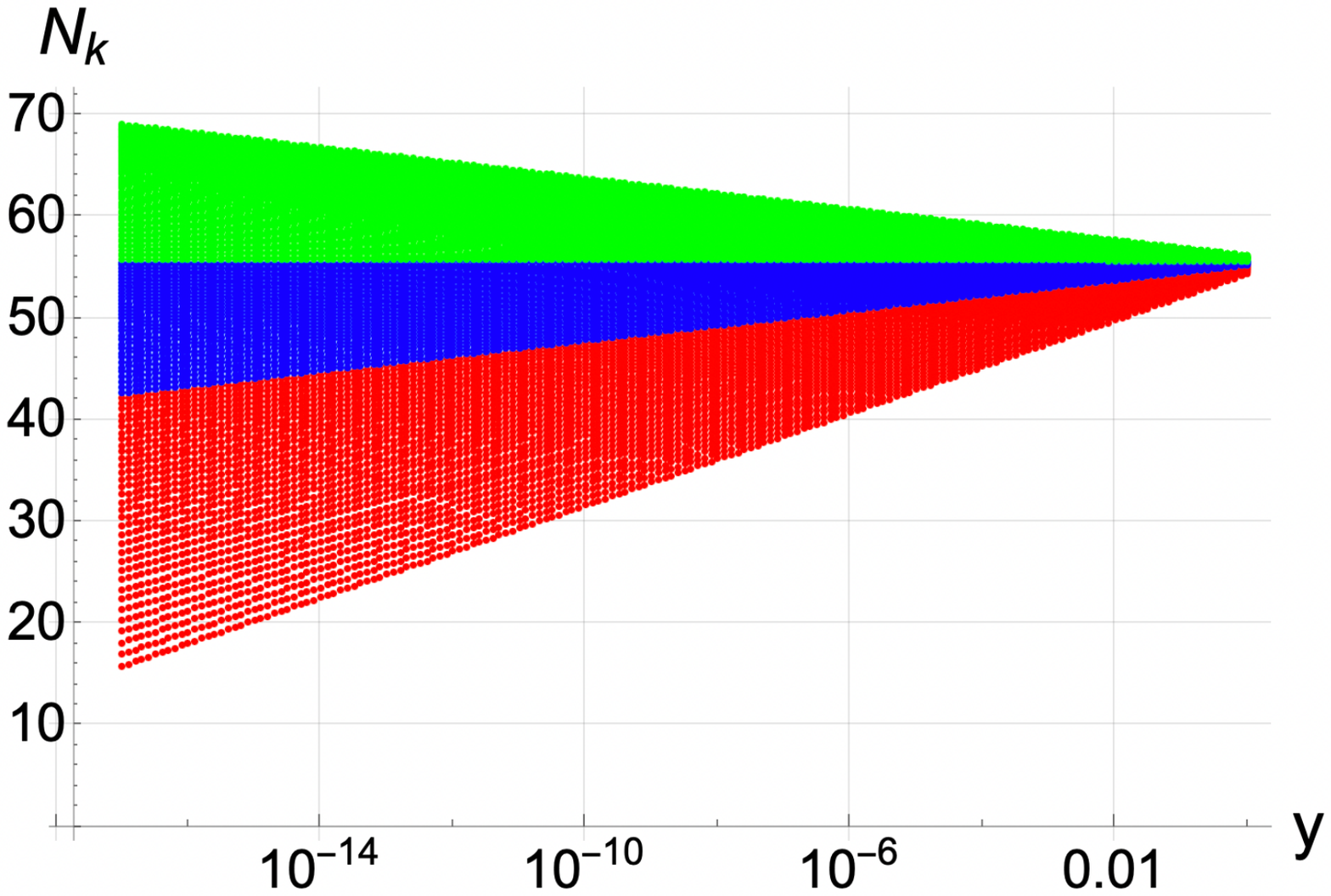}&
\includegraphics[width=3.in]{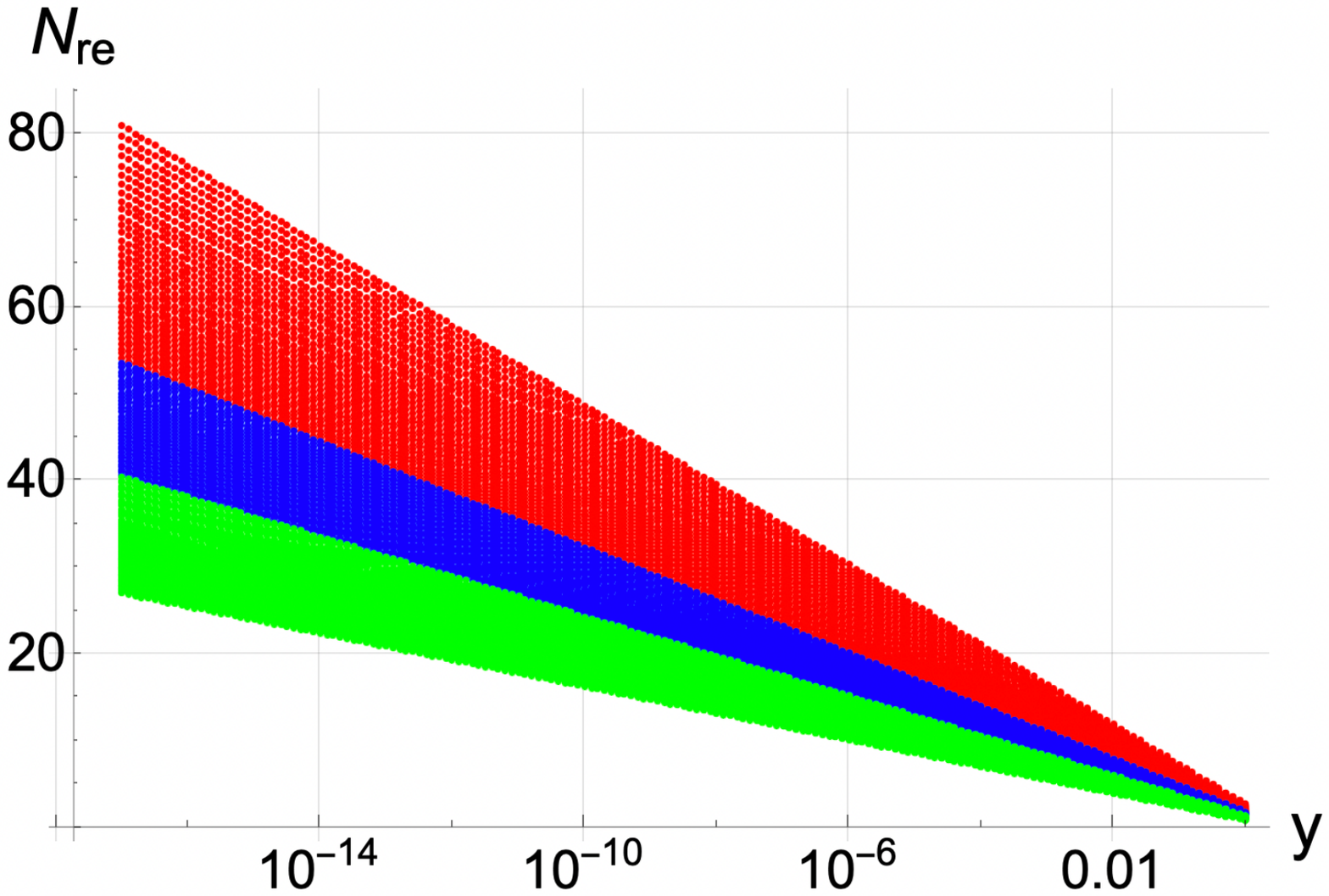}\\
\end{array}$
\caption{The plots show the number of $e$-folds during inflation $N_k$, Eq.~(\ref{Nk}), and during reheating $N_{re}$, Eq.~(\ref{Nre}), as functions of the coupling parameter $y$ (see Eq.~(\ref{Trey2})), for the $\alpha$-attractor model defined in Eq.~(\ref{pot}). The coupling $y$ is varied from $y \approx 1.71 \times 10^{-17}$ up to $y = 1$. The equation-of-state parameter $\omega_{re}$ is varied within the range $-1/3 < \omega_{re} < 1$, and divided into three intervals: red for $\omega_{re} \in [-1/3,\,0]$, blue for $\omega_{re} \in (0,\,1/3]$, and green for $\omega_{re} \in (1/3,\,1)$. Figures~\ref{A_Nk_y} to~\ref{A_nsk_y} are plotted for $\alpha = 1$. However, results for other values of $\alpha$ within the range $(0.001,\,37)$ are qualitatively and quantitatively similar.
}
\label{A_Nk_y}
\end{center}
\end{figure*}
\begin{figure*}[ht]
\begin{center}
$\begin{array}{ccc}
\includegraphics[width=3.in]{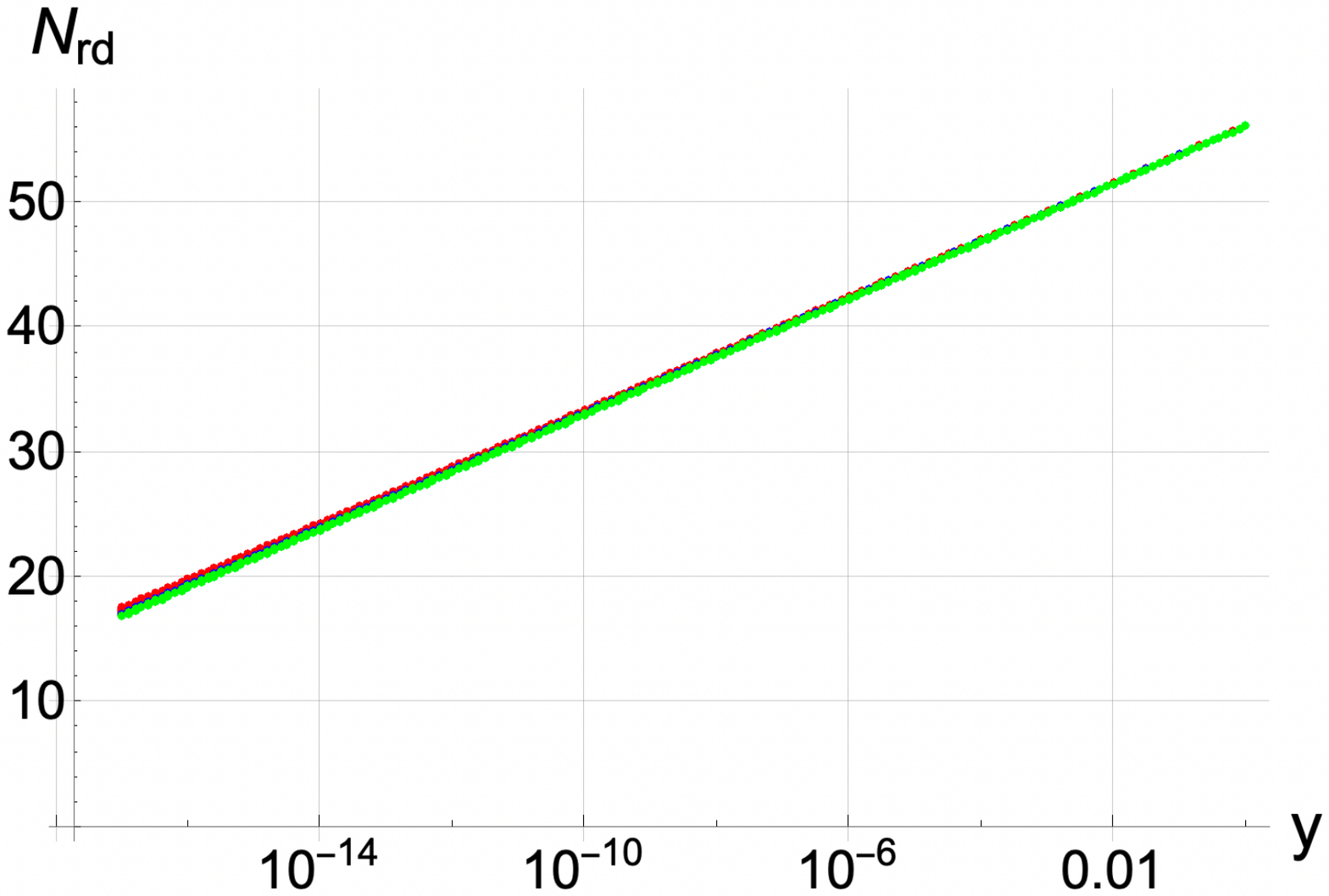}&
\includegraphics[width=3.in]{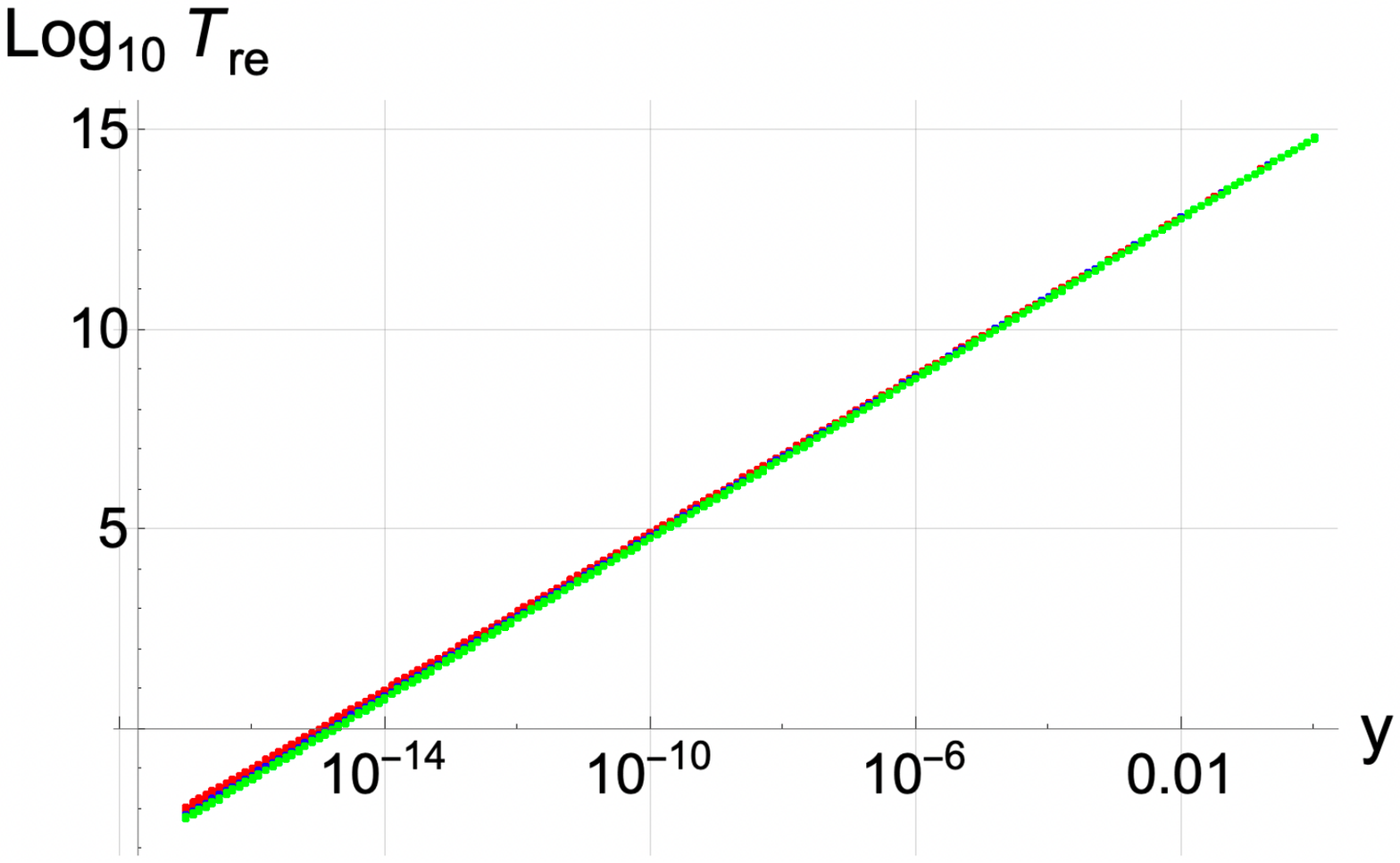}\\
\end{array}$
\caption{The plots show the number of $e$-folds during radiation $N_{rd} \equiv \ln\left(\frac{a_{eq}}{a_r}\right) = \ln\left(\frac{a_{eq} T_{re}}{\left(\frac{43}{11 g_{s,re}}\right)^{1/3} a_0 T_0}\right)$, and the reheating temperature $T_{re}$, given by Eq.~(\ref{Trey2}), as functions of the coupling parameter $y$, for the $\alpha$-attractor model defined in Eq.~(\ref{pot}). Other details are the same as in Fig.~\ref{A_Nk_y}.}
\label{A_Nrd_y}
\end{center}
\end{figure*}
\begin{figure*}[ht]
\begin{center}
\includegraphics[width=4.5 in]{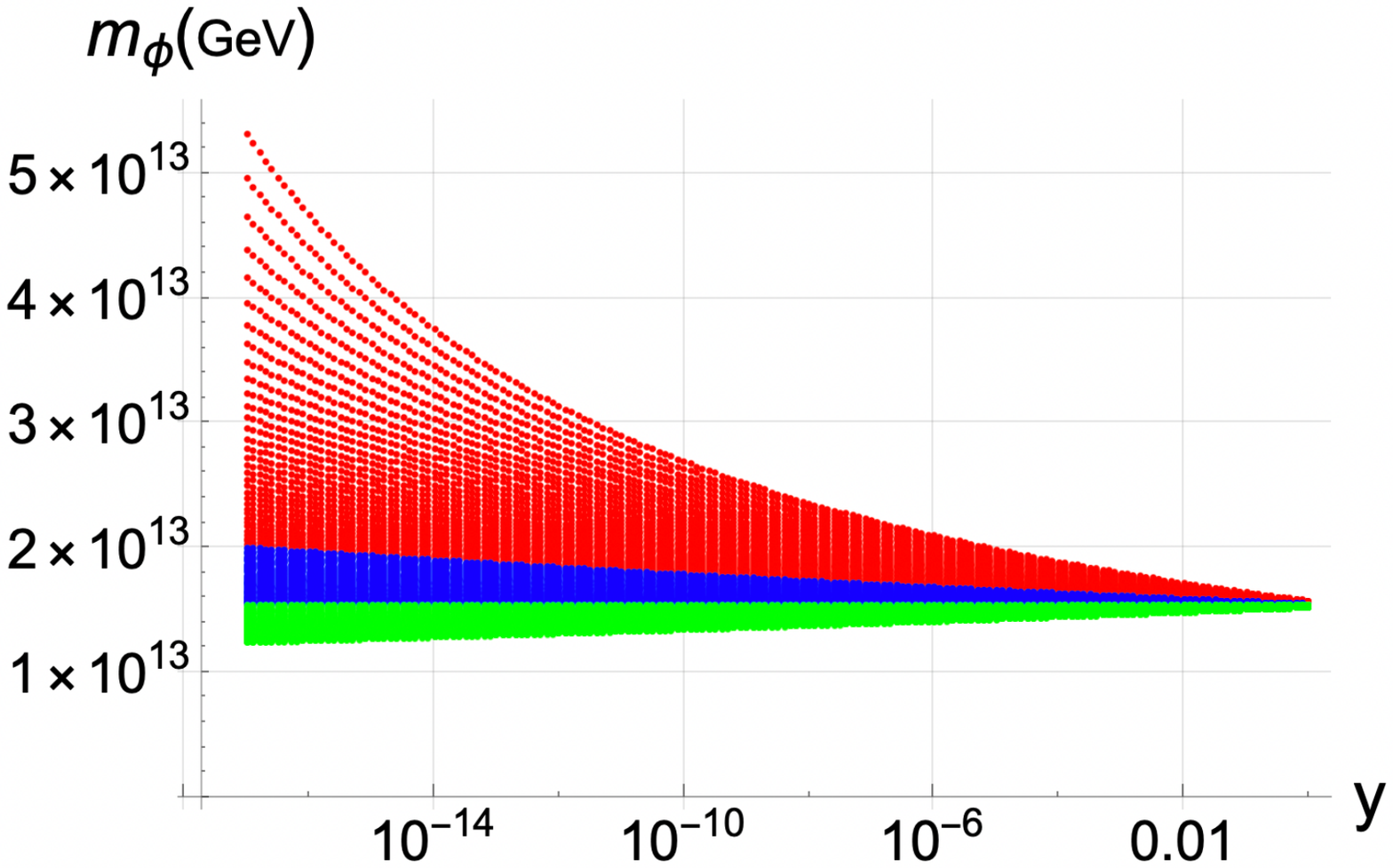}
\caption{The plot shows the inflaton mass $m_{\phi}$, given by Eq.~(\ref{mass2}), as a function of the coupling parameter $y$ (see Eq.~(\ref{Trey2})), for the $\alpha$-attractor model defined in Eq.~(\ref{pot}). Other details are the same as in Fig.~\ref{A_Nk_y}.}
\label{A_mfi_y}
\end{center}
\end{figure*}
\begin{figure*}[ht]
\begin{center}
$\begin{array}{ccc}
\includegraphics[width=3.in]{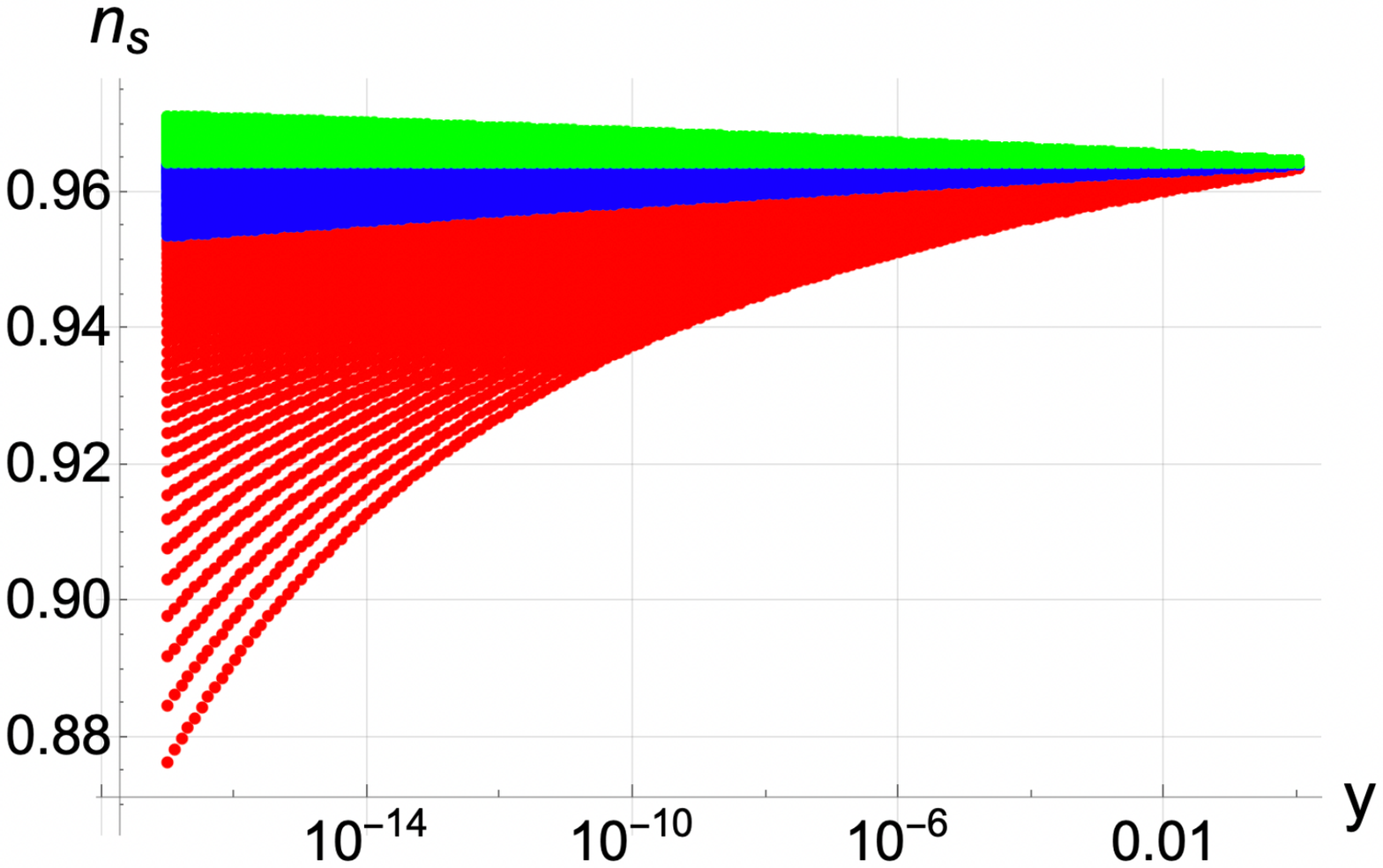}&
\includegraphics[width=3.in]{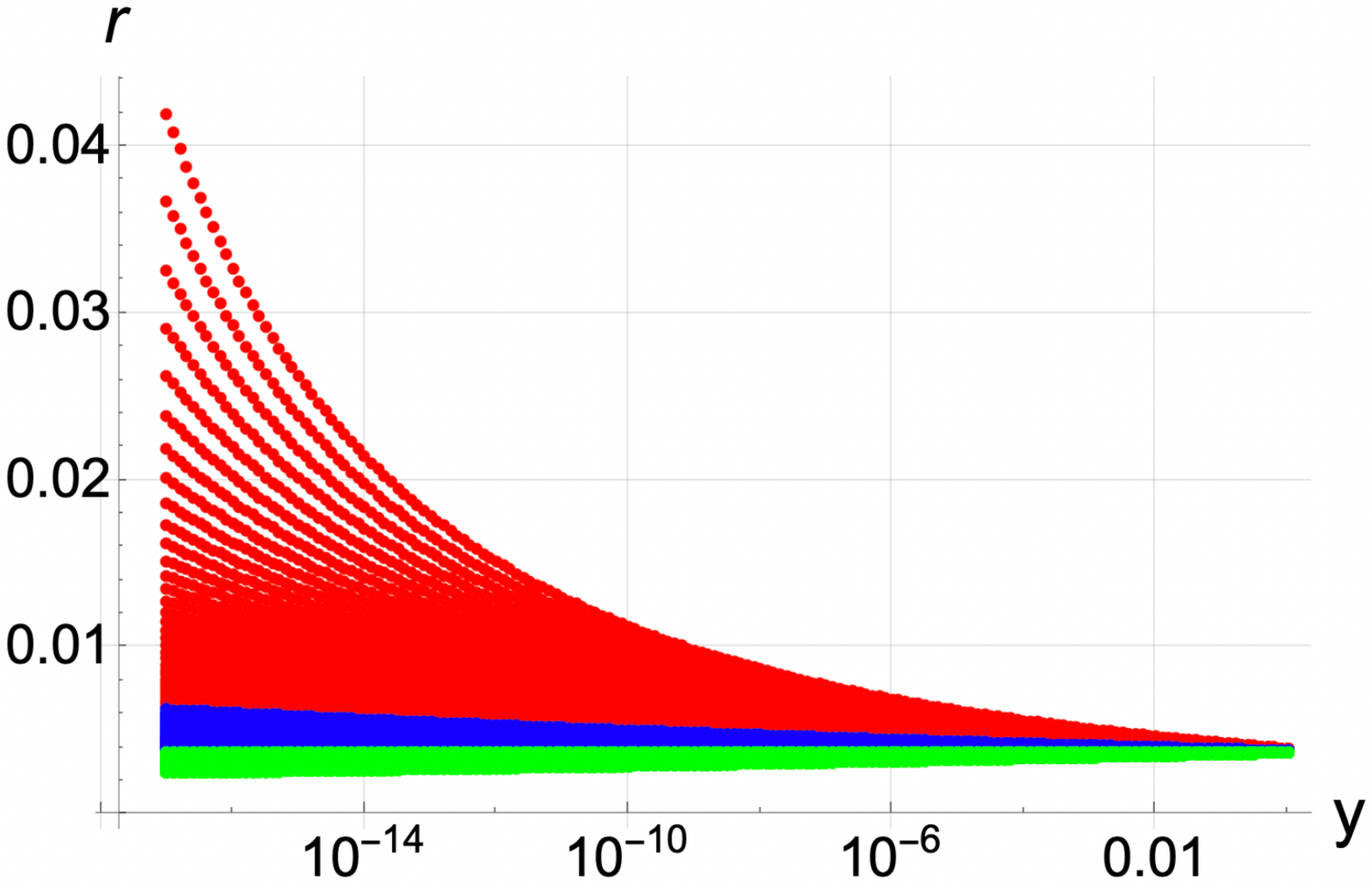}\\
\end{array}$
\caption{The plots show the scalar spectral index $n_s$, Eq.~(\ref{ns}), and the tensor-to-scalar ratio $r$, Eq.~(\ref{r}), as functions of the coupling parameter $y$ (see Eq.~(\ref{Trey2})), for the $\alpha$-attractor model defined in Eq.~(\ref{pot}). Other details are the same as in Fig.~\ref{A_Nk_y}.}
\label{A_ns_y}
\end{center}
\end{figure*}
\begin{figure*}[ht]
\begin{center}
\includegraphics[width=4.5 in]{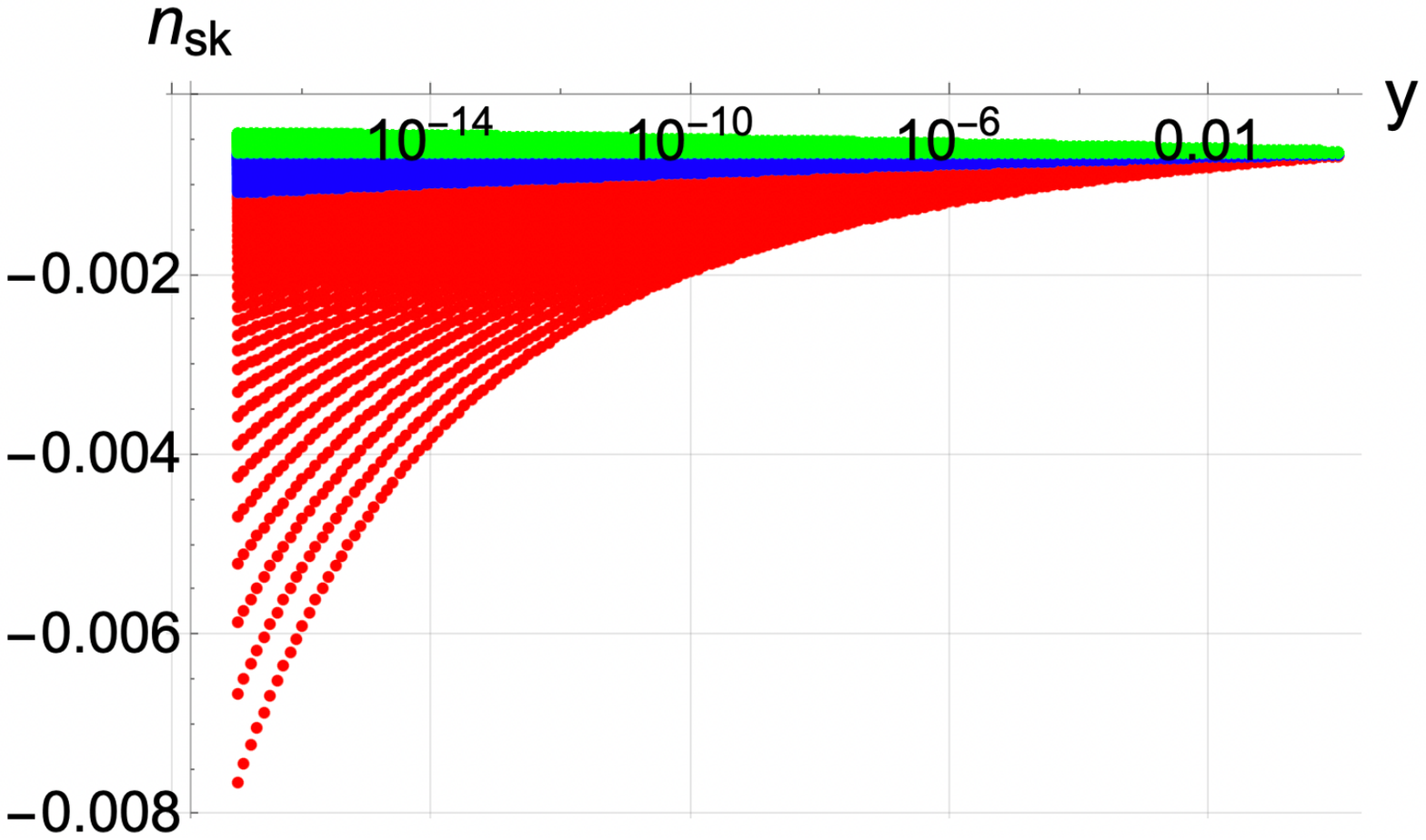}
\caption{The plot shows the running of the scalar spectral index $n_{sk}$, Eq.~(\ref{nsk}),  as a function of the coupling parameter $y$ (see Eq.~(\ref{Trey2})), for the $\alpha$-attractor model defined in Eq.~(\ref{pot}). Other details are the same as in Fig.~\ref{A_Nk_y}.}
\label{A_nsk_y}
\end{center}
\end{figure*}

\end{document}